\documentclass[lettersize,journal]{IEEEtran}
\usepackage{amsmath,amsfonts,amssymb}
\usepackage{algorithmic,algorithm}
\usepackage{array}
\usepackage[caption=false,font=normalsize,labelfont=sf,textfont=sf]{subfig}
\usepackage{textcomp}
\usepackage{stfloats}
\usepackage{url}
\usepackage{verbatim}
\usepackage{graphicx}
\def\BibTeX{{\rm B\kern-.05em{\sc i\kern-.025em b}\kern-.08em
    T\kern-.1667em\lower.7ex\hbox{E}\kern-.125emX}}
\usepackage{balance}
\usepackage{multirow}
\usepackage{xr}
\let\IEEEproof\proof
\let\IEEEendproof\endproof
\let\proof\relax
\let\endproof\relax
\usepackage{amsthm}
\let\proof\IEEEproof
\let\endproof\IEEEendproof

\theoremstyle{definition}
\newtheorem{remark}{Remark}
\newtheorem{proposition}{Proposition}
\newtheorem{lemma}{Lemma}
\newtheorem{corollary}{Corollary}
\newtheorem{theorem}{Theorem}

\begin{document}
\title{Adaptive Multi-kernel Correntropy Kalman Filter:
A Variational Bayesian Approach}
\author{Zhen Zeng, Shilei Li, \IEEEmembership{Member, IEEE}, Dawei Shi, \IEEEmembership{Senior Member, IEEE}, and Ling Shi, \IEEEmembership{Fellow, IEEE}
\thanks{Zhen Zeng is with the Department of Automation, Xiamen University, Xiamen, 361102, China (e-mail: Zengzhen200413@gmail.com).}
\thanks{Shilei Li is with the School of Automation, Beijing Institute of Technology, Beijing 100081, China (e-mail: shileili@bit.edu.cn).}
\thanks{Dawei Shi is with the School of Automation, Beijing Institute of Technology, Beijing 100081, China (e-mail: daweishi@bit.edu.cn).}
\thanks{Ling Shi is with the Department of Electronic and Computer Engineering, The Hong Kong University of Science and Technology, Hong Kong 999077, SAR (e-mail: eesling@ust.hk).}
}

\markboth{IEEE Transactions on Signal Processing, Vol.~XX, No.~XX, Month~20XX}%
{Adaptive Multi-kernel Correntropy Kalman Filter}

\maketitle

\begin{abstract}
This paper proposes an adaptive multi-kernel correntropy Kalman filter (AMCKF) for robust state estimation in the presence of heavy-tailed measurement noises and outliers. The multi-kernel correntropy loss is reformulated via half-quadratic (HQ) duality into a quadratic form with auxiliary weights, enabling efficient Kalman-like recursions. We establish a rigorous variational Bayesian framework under the mean-field assumption, deriving closed-form coordinate ascent variational inference updates for joint state and kernel width estimation. To enable online implementation, AMCKF is then proposed as an efficient single-pass approximation of the exact coordinate ascent variational inference (CAVI), where the state and auxiliary variables are updated iteratively via IRLS, followed by a closed-form kernel width update. A forgetting mechanism is designed to prevent over-confidence and ensure long-term adaptability. Simulations on target tracking and IMU attitude estimation demonstrate that the proposed AMCKF significantly outperforms standard KF, VBKF, Student's t-based KF, and fixed-kernel MKCKF in scenarios with sparse outliers, heavy-tailed noise, and mixed disturbances, while maintaining optimal performance under Gaussian conditions.
\end{abstract}


\section{Introduction}
\label{sec:intro}
\IEEEPARstart{S}{tate} estimation for noisy dynamical systems underpins modern signal processing and control, with the Kalman filter (KF) as the standard solution under linear-Gaussian assumptions~\cite{kalman1960, kalman1961, mehra1970identification}. The KF and its extensions have become ubiquitous across aerospace tracking, navigation, and robotics~\cite{ho1964bayesian, fitzgerald1971divergence, gusrial2025review}.

However, the optimality of the KF relies on precise knowledge of noise covariances. In practice, noise statistics are often unknown, time-varying, or subject to abrupt changes due to environmental disturbances, sensor degradation, or unmodeled dynamics~\cite{murartal2015, li2024a, xia2023}. Moreover, heavy-tailed measurement noise with outliers commonly exists in real sensors, severely degrading KF performance or even causing divergence~\cite{fitzgerald1971}.

Two main directions address these issues: adaptive filtering and robust filtering. Adaptive filtering estimates time-varying noise statistics online, with variational Bayesian (VB) methods gaining increasing attention. Särkkä and Nummenmaa pioneered a recursive VB framework for joint state and measurement noise covariance estimation~\cite{sarkka2009}, later extended to process noise~\cite{huang2018} and nonlinear systems~\cite{lan2023, lv2022, smidl2017variational}. Other adaptive approaches include covariance matching~\cite{meng2016}, innovation-based methods~\cite{mohamed1999}, and multiple-model strategies~\cite{li1994}. However, conventional VB filters are Gaussian-based and thus sensitive to outliers~\cite{he2021, chan2015robust}.

Robust filtering suppresses outliers via heavy-tailed models or robust cost functions. Student's \(t\)-distribution filters naturally downweight large errors and have been integrated into VB frameworks~\cite{rojas2019student, huang2017, agamennoni2012}. The maximum correntropy criterion (MCC) offers another effective solution~\cite{chen2017, liu2007, principe2010}. Compared with Student's \(t\) loss, the multi-kernel correntropy loss saturates more rapidly for large errors, providing stronger heavy-tailed noise suppression~\cite{li2021, li2024a, kulikova2020}. Consequently, MCC-based KFs exhibit strong robustness against impulsive noise~\cite{azimi2025stable}.

However, the practical performance of these robust filters depends critically on manually tuning the kernel width. To address this, several adaptive strategies have been proposed. Fakoorian et al.~\cite{fakoorian2019adaptive} proposed a heuristic weighted sum of the innovation and error covariance. Chen et al.~\cite{chen2023abmcckf} designed a protecting-window with chi-square detection to switch between MCC and MMSE. Li et al.~\cite{li2023vbckf} and Liu et al.~\cite{liu2023vbadaptive} explored variational Bayesian frameworks with Mahalanobis distance or resistance factors. Most recently, Zhang et al.~\cite{zhang2026improved} introduced an adaptive Cauchy kernel with an optimal factor to handle both outliers and state mutations for satellite attitude estimation.
Despite these advances, existing methods still rely on heuristic rules, discrete switching thresholds, or empirically tuned parameters, lacking a unified probabilistic framework for principled online kernel width learning. This motivates us to develop a variational Bayesian approach that learns the kernel width online.

To address these limitations, we establish a rigorous variational Bayesian framework with closed-form CAVI updates for joint state and kernel width estimation. The non-convex correntropy loss is reformulated via half-quadratic (HQ) duality~\cite{geman1992, rockafellar1970, nikolova2005} into a quadratic form with auxiliary weights, enabling Gaussian state updates. To enable online implementation, we then propose the adaptive multi-kernel correntropy Kalman filter (AMCKF) as an efficient single-pass approximation of the exact CAVI, where the state and auxiliary variables are updated iteratively via IRLS, followed by a closed-form update of the kernel widths. A forgetting mechanism prevents kernel width collapse under persistent outliers. The proposed framework extends naturally to a broader class of robust loss functions satisfying the HQ conditions.
The main contributions are as follows:
\begin{enumerate}
    \item We establish a general HQ duality result (\textbf{Proposition}~\ref{thm:linear_scaling}) showing that any loss of a certain form admits a decomposition with the scale parameter appearing linearly in the dual potential, unifying a family of robust losses.
    \item We develop a rigorous variational Bayesian framework under the mean-field assumption, deriving closed-form CAVI updates for joint state and kernel width estimation, and prove Gamma conjugacy (\textbf{Theorem}~\ref{prop:gamma_conjugacy}) for the HQ-induced likelihood.
    \item We propose AMCKF as an efficient single-pass approximation of the exact CAVI, where the state and auxiliary variables are updated iteratively via IRLS, followed by a closed-form kernel width update.
    \item Monte Carlo simulations demonstrate that the proposed AMCKF consistently outperforms KF, VBKF, Student's \(t\)-based KF, and fixed-kernel MKCKF under various heavy-tailed noise and outlier scenarios, while matching the accuracy of the CAVI algorithm.
\end{enumerate}

The remainder of this paper is organized as follows. Section~\ref{sec:problem} formulates the problem. Section~\ref{sec:main} presents the rigorous VB framework and the proposed AMCKF as its efficient approximation. Section~\ref{sec:simulation} presents simulation results. Section~\ref{sec:conclusion} concludes the paper.

\emph{Notations}: The \(n\times n\) identity matrix is denoted by \(I_n\). The transpose and inverse of a matrix \(A\) are denoted by \(A^T\) and \(A^{-1}\), respectively. \(\mathbb{E}[\cdot]\) denotes the expectation operator. The Gaussian distribution with mean \(\mu\) and covariance \(\Sigma\) is denoted by \(\mathcal{N}(\mu,\Sigma)\), and \(\Gamma(\alpha,\beta)\) denotes the Gamma distribution with shape \(\alpha\) and rate \(\beta\). The diagonal matrix with the vector \(v\) on its diagonal is denoted by \(\mathrm{diag}(v)\). The \(i\)-th component of a vector is denoted by \([\cdot]_i\).

\section{Problem Formulation}
\label{sec:problem}

Consider the following discrete-time linear dynamical system:
\begin{align}
    x_{k+1} &= A x_k + w_k, \label{eq:state} \\
    y_k &= C x_k + v_k, \label{eq:measurement}
\end{align}
where \(x_k \in \mathbb{R}^n\) is the system state,
\(y_k \in \mathbb{R}^m\) is the measurement vector,
\(w_k\) and \(v_k\) denote process and measurement noises.
Matrices \(A, B\) are system parameters,
and the pair \((A, C)\) is observable.
The initial state \(x_0\) follows a Gaussian distribution
with known mean \(\hat{x}_{0}^+\) and covariance \(P_{0}^+\),
and is independent of \(w_k\) and \(v_k\).

In standard Kalman filtering,
process and measurement noises are assumed to be zero-mean Gaussian white noises
with known nominal covariances:
\begin{align}
    w_k &\sim \mathcal{N}(0, Q_k), \\
    v_k &\sim \mathcal{N}(0, R_k), 
\end{align}
where \(Q_k\) and \(R_k\) are positive (semi-)definite matrices.
In practice, noise statistics are often unknown, time-varying, or heavy-tailed~\cite{he2021}.

Define the prior residual and the measurement residual as
\begin{equation}
e_k^{\mathrm{pre}} = \hat{x}_k^- - x_k, \qquad e_k^{\mathrm{mea}} = y_k - C x_k, \label{eq:pre_residual}
\end{equation}
where \(\hat{x}_k^- = A \hat{x}_{k-1}^+\) is the predicted state. The augmented residual vector is then
\begin{equation}
e_k = \begin{bmatrix} e_k^{\mathrm{pre}} \\ e_k^{\mathrm{mea}} \end{bmatrix}
    = \begin{bmatrix} \hat{x}_k^- - x_k \\ y_k - C x_k \end{bmatrix}
    = z_k - H_k x_k, \label{eq:augmented_residual}
\end{equation}
with
\begin{equation}
z_k = \begin{bmatrix} \hat{x}_k^- \\ y_k \end{bmatrix}, \qquad H_k = \begin{bmatrix} I \\ C \end{bmatrix}. \label{eq:zk_Hk}
\end{equation}
The covariance of the augmented residual is
\begin{equation}
\Sigma_k = \mathrm{Cov}(e_k) = \begin{bmatrix} P_k^- & 0 \\ 0 & R_k \end{bmatrix}, \label{eq:covariance}
\end{equation}
where \(P_k^-\) is the covariance of \(\hat{x}_k^- - x_k\) and \(R_k\) is the measurement noise covariance. Since \(\Sigma_k\) is block diagonal, its inverse square root is given by
\begin{equation}
\Sigma_k^{-1/2} = \begin{bmatrix} (P_k^-)^{-1/2} & 0 \\ 0 & R_k^{-1/2} \end{bmatrix}, \label{eq:sqrt_covariance}
\end{equation}
where \((P_k^-)^{-1/2}\) and \(R^{-1/2}\) are the square roots (e.g., Cholesky factors) of the respective inverse covariance matrices. The whitened residual is then defined as
\begin{equation}
\tilde{e}_k = \Sigma_k^{-1/2} e_k = \Sigma_k^{-1/2}(z_k - H_k x_k)=\tilde{z}_k-\tilde{H_k}x_k. \label{eq:whitened_residual}
\end{equation}

\begin{lemma}[Half-Quadratic Duality]
\label{lem:hq_duality}
Let \(\rho(e)\) be a robust loss function such that \(\phi(u) = \rho(\sqrt{u})\) is concave, continuously differentiable, and monotonically increasing on \(u \in [0, \infty)\). Then there exists a dual potential function \(\psi(w)\) such that:
\[
    \rho(e) = \inf_{w > 0} \left[ w\frac{e^2}{2} + \psi(w) \right].
\]
The infimum is achieved at \(w = \rho'(e) / e\).
\end{lemma}
For a detailed proof, see \cite{geman1992, rockafellar1970}.
Half-quadratic (HQ) duality has been widely 
used to reformulate non-convex loss functions into a form amenable to iterative reweighted least squares (IRLS) algorithms.

\begin{proposition}[Linear Scaling in Half-Quadratic Decomposition]
\label{thm:linear_scaling}
Let \(\rho(e;\theta)=\theta\,\Phi\!\left(\frac{e}{\sqrt{\theta}}\right)\) be a robust loss function, where \(\Phi\) satisfies the half-quadratic duality conditions (Lemma~\ref{lem:hq_duality}). Then the half-quadratic decomposition of \(\rho(e;\theta)\) takes the form
\[
\rho(e;\theta)=\min_{w>0}\left[ w\frac{e^{2}}{2}+\theta\,\psi(w)\right],
\]
where \(\psi(w)\) is the dual potential function obtained from the decomposition of \(\Phi\), i.e., \(\Phi(u)=\min_{w>0}\bigl[w u^{2}/2+\psi(w)\bigr]\). Consequently, \(\psi(w)\) does not depend on \(\theta\), and the scale parameter \(\theta\) appears linearly in the dual potential.
\end{proposition}

\begin{remark}
The significance of this proposition lies in its role as the structural prerequisite for Theorem~\ref{prop:gamma_conjugacy}. The linear appearance of \(\theta\) in the dual potential is precisely the condition that guarantees conjugacy of the Gamma prior in the variational Bayesian update.
\end{remark}

The proof is provided in the Appendix~A of the supplementary material.
The above proposition applies to a family of robust loss functions, including the cauchy loss, the charbonnier (pseudo-Huber) loss, fair loss and the multi-kernel correntropy loss, see details also in Appendix~A. 

Among these, the multi-kernel correntropy loss offers an explicit robustness–accuracy trade-off via its kernel width and will be the focus of this work.

\subsection{Maximum Correntropy Criterion and Multi-kernel Correntropy Loss} %强调MCKL跟其他鲁棒损失函数的不同,给出数学性质

Let \(e = [e_1, \dots, e_N]^\top\) be a random vector. The multi-kernel correntropy loss (MCKL) is defined element-wise as:
\begin{equation}
    \rho(e_i) = \sigma_i^2 \left(1 - \exp\left(-\frac{e_i^2}{2\sigma_i^2}\right)\right), \label{eq:mkcl}
\end{equation}
where \(\sigma_i^2 > 0\) is the kernel width for the \(i\)-th component.

The maximum correntropy criterion (MCC) maximizes the correntropy between true and estimated variables. For additive-error estimation tasks, minimizing the MCKL is equivalent to maximizing the MCC. Up to a scaling factor, \(\rho(e_i)\) coincides with the MCC loss:
\begin{equation}
    J_{\text{MCC}}(e) = \sum_{i=1}^{N} \left(1 - \exp\left(-\frac{e_i^2}{2\sigma_i^2}\right)\right) = \frac{1}{\sigma_i^2} \rho(e_i).
\end{equation}
Key properties of MCKL, including its boundedness, redescending influence function, and asymptotic behavior, are summarized in the supplementary materials.
Unlike Huber (non-redescending) and Student-\(t\) (polynomial decay), MKCL has exponential redescending influence and boundedness, with a single parameter \(\sigma^2\) controlling robustness.

\subsection{Half-Quadratic Decomposition of the Multi-kernel Correntropy Loss}

Recall that the MCKL is defined in \eqref{eq:mkcl}, we have:

\begin{corollary}
\label{prop:hq_decomposition}
The MKCL satisfies the conditions of the half-quadratic duality theorem, admitting the following exact decomposition:
\begin{equation}
    \rho(e_i) = \min_{w_i \in (0,1]} \left[ w_i \frac{e_i^2}{2} + \sigma_i^2 (w_i \log w_i - w_i + 1) \right].\label{eq:hq_mkcl}
\end{equation}
The unique minimizer is given by:
\begin{equation}
    w_i = \exp\left(-\frac{e_i^2}{2\sigma_i^2}\right), \label{eq:optimal_weight}
\end{equation}
\end{corollary}
The proof of this corollary is provided in the 
Appendix~B of the supplementary material.

\begin{remark}[Probabilistic Interpretation]
\label{rem:hq_probabilistic}
The HQ duality established in Lemma~\ref{lem:hq_duality} admits a direct probabilistic interpretation. For the auxiliary variable \(w_i\), which is introduced purely for optimization and for which no prior information is available, a uniform prior \(p(w_i) \propto \text{const}\) on \((0,1]\) is a natural non-informative choice. With the HQ-induced likelihood defined as
\begin{equation}
p(e_i \mid w_i, \sigma_i^2) \propto \exp\left( - w_i \frac{e_i^2}{2} - \sigma_i^2 \psi(w_i) \right),
\end{equation}
then the posterior of \(w_i\) satisfies
\begin{equation}
p(w_i \mid e_i, \sigma_i^2) \propto \exp\left( - w_i \frac{e_i^2}{2} - \sigma_i^2 \psi(w_i) \right).
\end{equation}
Maximizing this posterior yields
\begin{equation}
\hat{w}_i^{\text{MAP}} = \arg\min_{w_i} \left[ w_i \frac{e_i^2}{2} + \sigma_i^2 \psi(w_i) \right],
\end{equation}
which is precisely the HQ minimization in Lemma~\ref{lem:hq_duality}. So, the HQ minimizer is the MAP estimate for auxiliary variable \(w_i\). For the MCKL, Corollary~\ref{prop:hq_decomposition} gives:
\begin{equation}
\hat{w}_i^{\text{MAP}} = w_i^* = \exp\left(-\frac{e_i^2}{2\sigma_i^2}\right).
\end{equation}
\end{remark}

\subsection{HQ-induced unnormalizable Generalized Likelihood}

Apply the MKCL to the whitened residual \(\tilde{e}_k\) defined in \eqref{eq:whitened_residual}. Using the HQ decomposition \eqref{eq:hq_mkcl}, the combined loss function can be written as
\begin{equation}
\min_{x_k,\{w_{i,k}\}} \sum_{i=1}^{n+m} \left[ w_{i,k} \frac{[\tilde{e}_k]_i^2}{2} + \sigma_i^2 (w_{i,k}\log w_{i,k} - w_{i,k} + 1) \right], \label{eq:hq_obj}
\end{equation}
where the first \(n\) components correspond to the prediction error and the remaining \(m\) components to the measurement innovation. 
Let \(W = \operatorname{diag}(w_1,\dots,w_{n+m})\) be the weight matrix. Then \eqref{eq:hq_obj} can be compactly expressed as
\begin{equation}
\min_{x_k,\{w_{i,k}\}} \left[ \frac{1}{2} \tilde{e}_k^\top W \tilde{e}_k + \sum_{i=1}^{n+m} \sigma_i^2 (w_{i,k}\log w_{i,k} - w_{i,k} + 1) \right]. \label{eq:hq_obj_matrix}
\end{equation}
The term \(-\frac{1}{2} \tilde{e}_k^\top W \tilde{e}_k\) is quadratic in \(\tilde{e}_k\), and since \(\tilde{e}_k\) is affine in \(x_k\), its exponential defines a generalized likelihood:
\begin{equation}
\begin{aligned}
    &p(\tilde{e}_k\mid\{w_{i,k}\},\{\sigma_{i,k}^2\}) = p(y_k,\hat{x}_k^-\mid x_k,\{w_{i,k}\},\{\sigma_{i,k}^2\}) 
    \propto \\ &\exp\Bigg( 
    -\frac{1}{2} \tilde{e}_k^\top W \tilde{e}_k - \sum_{i=1}^{n+m} \sigma_{i,k}^2 (w_{i,k}\log w_{i,k} - w_{i,k} + 1)
    \Bigg).
\end{aligned}
\label{eq:generalized_likelihood_full}
\end{equation}
Minimizing \eqref{eq:hq_obj_matrix} is to maximize this likelihood over \(x_k\) and \(\{w_{i,k}\}\) with fixed \(\sigma_{i,k}^2\), which is equivalent to a conditional MAP problem.

\begin{remark}
The generalized likelihood in \eqref{eq:generalized_likelihood_full} implicitly incorporates the prior distribution of \(x_k\). It follows from the definition of the whitened residual that
\[
p(y_k, \hat{x}_k^- \mid x_k) = p(y_k, x_k \mid \hat{x}_k^-) \propto \exp\left( -\frac{1}{2} \tilde{e}_k^\top W \tilde{e}_k \right).
\]
When \(W = I\), expanding \(\tilde{e}_k\) gives
\[
\begin{aligned}
-\frac{1}{2} \tilde{e}_k^\top \tilde{e}_k = -\frac{1}{2} (x_k - \hat{x}_k^-)^T (P_k^-)^{-1} (x_k - \hat{x}_k^-) \\ - \frac{1}{2} (y_k - C x_k)^T R^{-1} (y_k - C x_k).
\end{aligned}
\]
Thus, maximizing \(p(y_k, \hat{x}_k^- \mid x_k)\) over \(x_k\) is equivalent to maximizing the posterior \(p(x_k \mid y_k, \hat{x}_k^-)=p(x_k \mid y_{1:k})\).
\end{remark}

\begin{remark}
The likelihood induced by the MKCL in \eqref{eq:mkcl} is unnormalizable: its tail decay is insufficient to yield a finite integral. The same property carries over to the HQ-decomposed form in \eqref{eq:generalized_likelihood_full}, since the HQ decomposition preserves the original loss function at its minimizer. This does not affect the subsequent variational Bayesian inference, which operates on the joint distribution rather than requiring a normalized likelihood; the resulting variational posteriors remain proper distributions.
\end{remark}

\subsection{Bayesian Formulation of the Estimation Problem}
\label{sec:bayesian_formulation}

Building on the probabilistic interpretation of the HQ decomposition (Remark~\ref{rem:hq_probabilistic}) and the generalized likelihood \eqref{eq:generalized_likelihood_full} derived above, we now formulate the joint estimation of \(x_k\) and the kernel widths \(\{\sigma_{i,k}^2\}\) in a Bayesian framework. The state prior is \(p(x_k)=\mathcal{N}(x_k;\hat{x}_k^-,P_k^-)\). For the kernel widths, we assign Gamma priors:
\begin{equation}
    p(\sigma_{i,k}^2) \sim \Gamma \left( \frac{\epsilon_1^{i,k}}{2} + 1, \frac{\epsilon_2^{i,k}}{2} \right), \label{eq:gamma_prior}
\end{equation}
where \(\epsilon_1^{i,k}, \epsilon_2^{i,k} > 0\) are hyperparameters of the conjugate Gamma prior. The logarithmic form of this prior is:
\begin{equation}
    \log p(\sigma_{i,k}^2) = \frac{\epsilon_1^{i,k}}{2} \log \sigma_{i,k}^2 - \frac{\epsilon_2^{i,k}}{2} \sigma_{i,k}^2 + \text{const}. \label{eq:log_prior}
\end{equation}
which are conjugate to the HQ-induced likelihood as established below.

\begin{theorem}[Gamma Conjugacy under Half-Quadratic Likelihood]
\label{prop:gamma_conjugacy}
Let \(\rho(e;\theta)\) be a robust loss function satisfying Proposition~\ref{thm:linear_scaling}, so that its half-quadratic decomposition takes the form
\[
\rho(e;\theta) = \min_{w>0}\left[ w\frac{e^2}{2} + \theta\,\psi(w) \right],
\]
where \(\psi(w)\) is free of \(\theta\). Consider the HQ-induced likelihood
\[
p(e\mid\theta) \propto \exp\left( -\frac{we^2}{2} - \theta\,\psi(w) \right)
\]
with \(\theta>0\) and fixed \(w\). Under a Gamma prior \(\Gamma(\epsilon_1/2+1,\epsilon_2/2)\) on \(\theta\), the posterior \(p(\theta\mid e)\) is also Gamma with updated hyperparameters
\[
\epsilon_1^* = \epsilon_1, \qquad \epsilon_2^* = \epsilon_2 + 2\psi(w).
\]
\end{theorem}
The proof is provided in the Appendix~C of the supplementary material.

Thus, the complete joint distribution of \(y_k\), \(x_k\), \(\{\sigma_{i,k}^2\}_{i=1}^{n+m}\), and \(\{w_{i,k}\}_{i=1}^{n+m}\) is:
\begin{equation}
    \begin{aligned}
    &p(y_k, x_k, \{\sigma_{i,k}^2\}, \{w_{i,k}\} \mid \hat{x}_k^-) \propto
    \\ &\exp\left(-\frac{1}{2} \tilde{e}_k^\top W \tilde{e}_k - \sum_{i=1}^{n+m} \sigma_{i,k}^2 \psi(w_{i,k})\right)\prod_{i=1}^{n+m} \Gamma\left(\sigma_{i,k}^2; \epsilon_1^{i,k}, \epsilon_2^{i,k}\right)
    \label{eq:joint_bayesian_general}
    \end{aligned}
\end{equation}

\subsection{Objectives}
\label{sec:objectives}

We aim to construct a recursive filtering algorithm that is adaptively robust to abnormal measurement errors. To this end, we require point estimates of the state \(x_k\) and the kernel widths \(\{\sigma_{i,k}^2\}_{i=1}^{n+m}\), obtained as the means of their respective posterior distributions:
\begin{equation}
\hat{x}_k \triangleq \mathbb{E}_{p(x_k \mid y_{1:k})}[x_k], \qquad 
\hat{\sigma}_{i,k}^2 \triangleq \mathbb{E}_{p(\sigma_{k}^2 \mid y_{1:k})}[\sigma_{i,k}^2].
\end{equation}
Direct computation from the joint distribution \eqref{eq:joint_bayesian_general} is intractable. We therefore adopt a mean-field variational approximation
\begin{equation}
q(x_k, \{\sigma_{i,k}^2\}, \{w_{i,k}\}) = q(x_k) \prod_{i=1}^{n+m} q(\sigma_{i,k}^2) q(w_{i,k}),
\label{eq:mean_field}
\end{equation}
and minimize its Kullback-Leibler divergence to the true posterior, equivalently maximizing the evidence lower bound. The posterior means \(\hat{x}_k\) and \(\hat{\sigma}_{i,k}^2\) are then computed from the resulting variational distributions and propagated via certain prediction equations, yielding a recursive filtering algorithm.

In this paper, we focus on adapting the kernel widths of the measurement channels (\(i=n+1,\dots,n+m\)); the process noise kernel widths (\(i=1,\dots,n\)) are fixed at sufficiently large values so that the corresponding weights \(w_i\) are effectively unity, in which case the process noise update reduces to that of the standard Kalman filter. The extension to fully adaptive process noise is straightforward.

\section{Main Results}
\label{sec:main}

We first derive an exact variational Bayesian inference algorithm from the joint distribution \eqref{eq:joint_bayesian_general}. Under the mean-field assumption, coordinate ascent yields closed-form CAVI updates for \(q(x_k)\), \(q(w_{i,k})\), and \(q(\sigma_{i,k}^2)\), summarized in Algorithm~\ref{alg:cavi_exact}. The exact algorithm requires iterative updates until ELBO convergence at each time step.

We then propose an efficient approximation for online implementation. Fixing \(\sigma_{i,k}^2\) to its current mean and performing a single CAVI pass reduces the update of \(q(x_k)\) and \(q(w_{i,k})\) to a conditional MAP problem solved by IRLS, yielding the AMCKF algorithm.

\subsection{Variational Bayesian Inference via CAVI}
\label{sec:cavi}

We now derive the exact variational Bayesian inference algorithm based on the joint distribution \eqref{eq:joint_bayesian_general} and the mean-field assumption \eqref{eq:mean_field}. The evidence lower bound (ELBO) is defined as
\begin{equation}
\mathcal{L}(q) = \mathbb{E}_q\left[\log \frac{p(y_k, x_k, \{\sigma_{i,k}^2\}, \{w_{i,k}\} \mid \hat{x}_k^-)}{q(x_k, \{\sigma_{i,k}^2\}, \{w_{i,k}\})}\right].
\end{equation}

A key result underpinning the subsequent derivations is the equivalence between MAP estimation and variational inference with a degenerate (Delta) variational distribution.

\begin{lemma}[Equivalence between Degenerate Variational Inference and MAP Estimation]
\label{lem:delta_map}
Let \(p(y,\theta,\zeta)\) be a joint distribution with observations \(y\), target variable \(\theta\), and other variables \(\zeta\), and let
\[
\mathcal{L}(q) = \mathbb{E}_q\left[\log \frac{p(y,\theta,\zeta)}{q(\theta,\zeta)}\right]
\]
denote the evidence lower bound (ELBO). Under the mean-field assumption \(q(\theta,\zeta)=q(\theta)q(\zeta)\), if \(q(\theta)\) is restricted to the family of Delta distributions, then maximizing the ELBO with respect to \(q(\theta)\) is equivalent to computing the conditional MAP estimate:
\[
\max_{q(\theta)=\delta(\theta-\hat{\theta})} \mathcal{L}(q)
\;\Longleftrightarrow\;
\hat{\theta} = \arg\max_{\theta}\, \mathbb{E}_{q(\zeta)}[\log p(y,\theta,\zeta)].
\]
\end{lemma}

\begin{corollary}[Variational Realization of HQ Minimization]
\label{cor:hq_variational}
Let \(w\) be an auxiliary variable introduced via the HQ decomposition of a robust loss \(\rho(e;\sigma^2)\), with uniform prior \(p(w)\propto\mathrm{const}\). Under the mean-field assumption, restricting the variational distribution of \(w\) to the Delta family
\[
q(w) = \delta(w - \tilde{w})
\]
implies, by Lemma~\ref{lem:delta_map}, that maximizing the ELBO with respect to \(\tilde{w}\) is equivalent to MAP estimation. Combined with the probabilistic interpretation of the HQ decomposition (Remark~\ref{rem:hq_probabilistic}), this MAP estimate coincides with the HQ minimizer:
\[
\tilde{w} = \arg\min_{w} \left[ w\frac{e^2}{2} + \sigma^2 \psi(w) \right].
\]
Hence, the variational optimization of \(w\) simultaneously accomplishes the minimization required by the HQ duality in Lemma~\ref{lem:hq_duality}.
\end{corollary}

We now specify the variational form of the auxiliary variables \(\{w_{i,k}\}_{i=1}^{n+m}\). By Corollary~\ref{cor:hq_variational}, for each \(w_{i,k}\), the optimal variational distribution is degenerate:
\begin{equation}
q(w_{i,k}) = \delta(w_{i,k} - \tilde{w}_{i,k}).
\end{equation}
The HQ decomposition is thus naturally embedded into the variational Bayesian framework. The coordinate ascent updates for the remaining factors are derived below.

\paragraph{Update of \(q(w_{i,k})\)}
With \(q(w_{i,k})=\delta(w_{i,k}-\tilde{w}_{i,k})\), maximizing the ELBO with respect to \(\tilde{w}_{i,k}\) via Lemma~\ref{lem:delta_map} yields
\begin{equation*}
\tilde{w}_{i,k} = \arg\min_{w_{i,k}} \left( w_{i,k} \frac{\mathbb{E}_{q(x_k)}[ [\tilde{e}_k(x_k)]_i^2 ]}{2} + \mathbb{E}_{q(\sigma_{k}^2)}[\sigma_{i,k}^2] \psi(w_{i,k}) \right)
\end{equation*}
where \(\psi(w_{i,k}) = w_{i,k} \log w_{i,k} - w_{i,k} + 1\). Solving gives
\begin{equation}
\tilde{w}_{i,k} = \exp\left( -\frac{\mathbb{E}_{q(x_k)}[[e_k(x_k)]_i^2]}{2 \mathbb{E}_{q(\sigma_k^2)}[\sigma_{i,k}^2]} \right), \label{eq:vb_update_w_i}
\end{equation}

\paragraph{Update of \(q(x_k)\)}
Fixing \(q(\sigma_k^2)\) and \(q(w_k)\), collecting all terms in the ELBO that depend on \(x_k\) yields
\begin{equation}
\log q(x_k) = -\frac{1}{2} \tilde{e}_k^\top \tilde{W}_k \tilde{e}_k + \text{const},
\end{equation}
where \(\tilde{W}_k = \operatorname{diag}(\tilde{w}_{1,k}, \dots, \tilde{w}_{n+m,k})\) with \(\tilde{w}_{i,k} = \mathbb{E}_{q(w_k)}[w_{i,k}]\). This quadratic form implies that \(q(x_k)\) is Gaussian:
\begin{equation*}
q(x_k) = \mathcal{N}(x_k; \mu_k, P_k^+)
\end{equation*}
with
\begin{align}
P_k^+ &= \left( \tilde{H}_k^\top \tilde{W}_k \tilde{H}_k \right)^{-1}\\
\mu_k &= P_k^+ \tilde{H}_k^\top \tilde{W}_k \tilde{z}_k.
\end{align}

With \(q(x_k)\) now determined, the expectation appearing in \eqref{eq:vb_update_w_i} can be evaluated explicitly:
\begin{equation}
\mathbb{E}_{q(x_k)}[ [e_k(x_k)]_i^2 ] = [e_k(\mu_k)]_i^2 + [H_k P_k^+ H_k^\top]_{ii}. \label{eq:expand_var}
\end{equation}

\begin{remark}[Weight Structure under Fixed Process Noise Kernels]
\label{rem:cavi_weight_structure}
By fixing the process noise kernel widths to sufficiently large values, the corresponding weights satisfy \(\tilde{w}_i = 1\) for \(i=1,\dots,n\). The weight matrix thus becomes block-diagonal:
\[
\bar{W}_k = \begin{bmatrix} I_n & 0 \\ 0 & \bar{W}_{\rm meas,k} \end{bmatrix}, 
\bar{W}_{\rm meas,k} = \operatorname{diag}(\bar{w}_{n+1,k}, \dots, \bar{w}_{n+m,k}).
\]
Substituting this into the expressions for \(\mu_k\) and \(P_k^+\) with
\[
\tilde{z}_k = \begin{bmatrix} (P_k^-)^{-1/2} \hat{x}_k^- \\ R_k^{-1/2} y_k \end{bmatrix}, \qquad
\tilde{H}_k = \begin{bmatrix} (P_k^-)^{-1/2} \\ R_k^{-1/2} C \end{bmatrix}
\]
gives
\begin{align}
P_k^+ &= \left( (P_k^-)^{-1} + C^\top R^{-1/2}_k \bar{W}_{\rm meas,k} R^{-1/2}_k C \right)^{-1},\label{eq:P_update} \\
\mu_k &= P_k^+ \left( (P_k^-)^{-1} \hat{x}_k^- + C^\top R^{-1/2}_k \bar{W}_{\rm meas,k} R^{-1/2}_k y_k \right).\label{eq:mu_update}
\end{align}
Thus, the process noise channel retains the standard Kalman filter structure, while the measurement channels undergo adaptive robust weighting.
\end{remark}

\paragraph{Update of \(q(\sigma_{i,k}^2)\)}
The following corollary, which follows directly from Theorem~\ref{prop:gamma_conjugacy}, establishes the conjugacy for the specific loss used in this work.

\begin{corollary}
\label{cor:gamma_conjugacy}
For the multi-kernel correntropy loss, Proposition~\ref{thm:linear_scaling} applies with \(\psi(w_{i,k})=w_{i,k}\log w_{i,k}-w_{i,k}+1\). Hence, according to Theorem~\ref{prop:gamma_conjugacy}, under the Gamma prior, the posterior of \(\sigma_{i,k}^2\) is also Gamma with updated hyperparameters
\begin{align}
    \epsilon_1^{i,k,+} &= \epsilon_1^{i,k},\label{eq:epsilon1_update} \\
    \epsilon_2^{i,k,+} &= \epsilon_2^{i,k} + 2(w_{i,k}\log w_{i,k} - w_{i,k} + 1)
    \label{eq:epsilon2_update}.
\end{align}
\end{corollary}

Fixing \(q(x_k)\) and \(q(w_{i,k})\), then by Corollary~\ref{cor:gamma_conjugacy}, the optimal \(q(\sigma_i^2)\) is
\begin{equation}
q(\sigma_{i,k}^2) = \Gamma\left(\sigma_{i,k}^2; \frac{\epsilon_1^{i,k,+}}{2}+1, \frac{\epsilon_2^{i,k,+}}{2}\right),
\end{equation}
with
\begin{equation}
\epsilon_2^{i,k,+} = \epsilon_2^{i,k} + 2(\tilde{w}_{i,k} \log \tilde{w}_{i,k} - \tilde{w}_{i,k} + 1).\label{eq:epsilon2_update_cavi}
\end{equation}
Its posterior mean is
\begin{equation}
\mathbb{E}_{q(\sigma_k^2)}[\sigma_{i,k}^2] = \frac{\epsilon_1^{i,k,+} + 2}{\epsilon_2^{i,k,+}}. \label{eq:sigma_posterior_mean}
\end{equation}

\begin{remark}[Active Set]
    In practice, some measurement channels may be known to be reliable and do not require adaptive robustness. Therefore, we introduce an \emph{active set} \(\mathcal{A} \subseteq \mathcal{I}\), where \(\mathcal{I}\) denotes the set of all measurement channels. Only the kernel widths \(\sigma_{i,k}^2\) for \(i \in \mathcal{A}\) are adaptively estimated; for \(i \notin \mathcal{A}\), \(\sigma_{i}^2\) is fixed at a nominal value.
\end{remark}

The above updates constitute a CAVI scheme: starting from an initial guess of the variational parameters, \(q(w_i)\), \(q(x_k)\), and \(q(\sigma_i^2)\) are updated iteratively in a coordinate-wise manner until convergence of the ELBO.

\subsection{Algorithmic Flow}
\label{sec:algorithmic_flow}

We now present the complete recursive filtering algorithm based on the CAVI updates derived in Section~\ref{sec:cavi}. At each time step \(k\), given the prior \(\hat{x}_k^-\), \(P_k^-\), and hyperparameters \(\epsilon_1^{i,k}\), \(\epsilon_2^{i,k}\), the CAVI procedure is executed to approximate the joint posterior.

\paragraph{Initialization}
The CAVI iterations are initialized by setting the auxiliary weights to
\begin{equation}
\tilde{w}_{i,k}^{(0)} = 1, \quad i = 1,\dots,n+m.
\end{equation}
The remaining variational parameters are then obtained by sequentially applying the CAVI updates derived in Section~\ref{sec:cavi}: with \(\tilde{w}_i^{(0)} = 1\), the update of \(q(x_k)\) yields \(\mu_k^{(0)} = \hat{x}_k^-\) and \(P_k^{(0)} = P_k^-\); subsequently, the update of \(q(\sigma_{i,k}^2)\) yields \(\epsilon_2^{i,k,(0)}=\epsilon_2^{i,k}\), hence yields the prior mean \(\hat{\sigma}_i^{2,(0)} = (\epsilon_1^{i,k} + 2)/\epsilon_2^{i,k,(0)}\).

\paragraph{CAVI iteration}
At each iteration \(t\), the updates are performed sequentially:
\begin{enumerate}
    \item Update \(q(w_i)\): compute \(\tilde{w}_i^{(t+1)}\) via \eqref{eq:vb_update_w_i} using \(\mu_k^{(t)}\), \(P_k^{(t)}\), and \(\hat{\sigma}_i^{2,(t)}\);
    \item Update \(q(x_k)\): compute \(\mu_k^{(t+1)}\), \(P_k^{(t+1)}\) via \eqref{eq:P_update} and \eqref{eq:mu_update} using current \(\tilde{w}_i^{(t+1)}\);
    \item Update \(q(\sigma_i^2)\): compute \(\epsilon_2^{i,k,(t+1)}\) via \eqref{eq:epsilon2_update_cavi} and \(\hat{\sigma}_i^{2,(t+1)}\) via \eqref{eq:sigma_posterior_mean} using \(\tilde{w}_i^{(t+1)}\).
\end{enumerate}
Iterations proceed until the ELBO converges.

\paragraph{Posterior extraction and propagation}
Upon convergence at iteration \(T_k\), the posterior means are extracted as
\begin{equation}
\hat{x}_k^+ = \mu_k^{(T_k)}, P_k^+ = P_k^{(T_k)}, \epsilon_2^{i,k,+} = \epsilon_2^{i,k,(T_k)}, \hat{\sigma}_{i,k}^{2,+} = \hat{\sigma}_i^{2,(T_k)}.
\end{equation}
The state estimate is propagated via the standard Kalman prediction:
\begin{equation}
\hat{x}_{k+1}^- = A \hat{x}_k^+, \qquad P_{k+1}^- = A P_k^+ A^\top + Q_k. \label{eq:prediction_step}
\end{equation}
The hyperparameters are propagated as
\begin{equation}
\epsilon_1^{i,k+1} = \epsilon_1^{i,k,+}, \qquad
\epsilon_2^{i,k+1} = \rho \, \epsilon_2^{i,k,+},\label{eq:hyperparameter_propagation}
\end{equation}
where \(\rho \in (0,1]\) is a forgetting factor.

\paragraph{Motivation for forgetting}
In normal operation, \(\tilde{w}_{i,k} \approx 1\) and \(\hat{\sigma}_{i,k}^{2,+}\) remains nominal. When an outlier occurs, \(\tilde{w}_{i,k}\) decreases, reducing \(\hat{\sigma}_{i,k}^{2,+}\) and down-weighting the corrupted measurement. However, without a forgetting mechanism, \(\hat{\sigma}_{i,k}^{2,+}\) is monotonically non-increasing and may collapse under prolonged outliers, causing over-confidence and mistaking genuine state changes for outliers. The forgetting factor \(\rho\) allows \(\hat{\sigma}_{i,k}^{2,+}\) to recover during normal operation, ensuring long-term adaptability.

\begin{remark}[Kernel Width Bounds]
To prevent the kernel width from becoming excessively large (which would forfeit robustness) or collapsing to zero (which would cause over-confidence), explicit bounds are imposed on the hyperparameter:
\begin{equation}
\epsilon_2^i \in [\epsilon_{2,\min}^i, \epsilon_{2,\max}^i].
\end{equation}
Since \(\hat{\sigma}_i^2 = (\epsilon_1^i + 2)/\epsilon_2^i\), this directly bounds the kernel width. In practice, \(\epsilon_{2,\max}^i\) is typically set to \(10\) to \(100\) times \(\epsilon_{2,\min}^i\), depending on the desired robustness level.
\end{remark}

\subsection{EWMA Filtering for Hyperparameter Evolution}

The kernel width update based solely on instantaneous innovation may cause abrupt fluctuations of \(\hat{\sigma}_{i,k}^2\) under transient outliers. To leverage historical information, an exponential moving average (EWMA) filter can be applied to the hyperparameter update:
\begin{equation}
\epsilon_2^{i,k,+} = \beta \, \epsilon_2^{i,k-1,+} + (1-\beta) \, \epsilon_2^{i,k,(T_k)}
\label{eq:ewma_epsilon2}
\end{equation}
where \(\beta \in [0,1)\) is a filtering factor and the case \(\beta = 0\) recovers the instantaneous update.

This EWMA filtering provides a tunable trade-off between responsiveness and stability. It is optional and does not affect the core variational Bayesian derivation; historical information can be incorporated when smoother adaptation is desired.

Combining the CAVI updates, the recursive propagation, and the optional EWMA filtering, the complete rigorous variational Bayesian inference algorithm is summarized in Algorithm~\ref{alg:cavi_exact}.

\begin{algorithm}[!t]
\caption{CAVI for Joint State and Kernel Width Estimation}
\label{alg:cavi_exact}
\begin{algorithmic}[1]
\STATE \textbf{Input:} System matrices \(A, C\), noise covariance \(R_k,Q_k\), initial state \(\hat{x}_{0|0}\) and \(P_{0|0}\), initial hyperparameters \(\epsilon_1^{i,0}\), \(\epsilon_2^{i,0}\), forgetting factor \(\rho\), filtering factor \(\beta\)
\STATE \textbf{Output:} State estimates \(\hat{x}_k^+\) and covariance \(P_k^+\), kernel width estimates \(\hat{\sigma}_{i,k}^{2,+}\)
\FOR{each time step \(k = 1, 2, \dots\)}
    \STATE \textbf{Prediction:}
\STATE get \(\hat{x}_k^-\) and \(P_k^-\) via \eqref{eq:prediction_step}
\FOR{\(i \in \mathcal{A}\)}
    \STATE get \(\epsilon_1^{i,k}\), \(\epsilon_2^{i,k}\) via \eqref{eq:hyperparameter_propagation}
\ENDFOR
    \STATE \textbf{CAVI Initialization:} \(\tilde{w}_{i,k}^{(0)} = 1\), \(t = 0\)
    \REPEAT
        \STATE \(t \gets t + 1\)
        \STATE \textbf{Update \(q(w_i)\):} compute \(\tilde{w}_{i,k}^{(t)}\) via \eqref{eq:vb_update_w_i} using \(\mu_k^{(t-1)}\), \(P_k^{(t-1)}\), \(\hat{\sigma}_i^{2,(t-1)}\)
        \STATE \textbf{Update \(q(x_k)\):} compute \(\mu_k^{(t)}\), \(P_k^{(t)}\) via \eqref{eq:P_update} and \eqref{eq:mu_update} using \(\tilde{w}_{i,k}^{(t)}\)
        \FOR{\(i \in \mathcal{A}\)}
            \STATE \textbf{Update \(q(\sigma_i^2)\):} compute \(\epsilon_2^{i,k,(t)}\) via \eqref{eq:epsilon2_update_cavi} and \(\hat{\sigma}_i^{2,(t)}\) via \eqref{eq:sigma_posterior_mean} using \(\tilde{w}_{i,k}^{(t)}\)
        \ENDFOR
    \UNTIL{ELBO convergence}
    \STATE \textbf{Posterior Extraction:}
    \STATE \(\hat{x}_k^+ = \mu_k^{(t)}\), \(P_k^+ = P_k^{(t)}\)
    \FOR{\(i \in \mathcal{A}\)}
        \STATE get \(\epsilon_2^{i,k,+}\) via \eqref{eq:ewma_epsilon2}
        \STATE \textbf{Apply Bounds:} \(\epsilon_2^{i,k,+} \in [\epsilon_{2,\min}^i, \epsilon_{2,\max}^i]\)
        \STATE Compute kernel width: get \(\hat{\sigma}_{i,k}^{2,+}\) via \eqref{eq:sigma_posterior_mean}
    \ENDFOR
\ENDFOR
\end{algorithmic}
\end{algorithm}

\subsection{The AMCKF Algorithm}
\label{sec:amckf_derivation}

The rigorous CAVI algorithm derived in the preceding sections provides a principled variational Bayesian solution, but requires iterative coordinate updates until ELBO convergence at each time step. This computational burden motivates an efficient approximation for online applications.

We introduce two simplifications. First, we restrict \(q(x_k)\) to a Delta distribution during optimization, and after convergence to the mode, restore the posterior to a Gaussian form with covariance given by the Joseph update \eqref{eq:kf_covariance_update}, which provides a Laplace-like approximation to the true posterior that is exact only when residuals are small (\(e_i \ll \sigma_i^2\)); for large residuals, the true MCKL Hessian would require the additional factor \((1 - e_i^2/\sigma_i^2)\), in which case the Joseph update yields a more conservative (larger) covariance estimate than the true Laplace approximation. In contrast, the exact CAVI is known to produce over-confident (underestimated) covariance estimates for \(x_k\) under heavy-tailed noise; by recovering the covariance only after convergence, AMCKF mitigates this over-confidence. Second, we perform only one M-step per time step: the point estimates of \(x_k\) and \(w_i\) are obtained iteratively via IRLS until convergence (with \(\sigma_i^2\) fixed), followed by a single evaluation of the closed-form update for \(\sigma_i^2\) in \eqref{eq:epsilon2_update_cavi}, since the dominant coupling between \(x_k\) and \(w_i\) is captured in the IRLS procedure and the subsequent \(\sigma_i^2\) update is typically a small correction.

Under these approximations, the weight computation becomes
\begin{equation}
\tilde{w}_{i,k} = \exp\left( -\frac{[\tilde{e}_k(\hat{x}_k)]_i^2}{2 \hat{\sigma}_{i,k}^2} \right),
\end{equation}
and the joint variational bayesian estimation of \(x_k\) and \(w_i\) reduces to a conditional MAP problem (by Lemma~\ref{lem:delta_map}) with \(\sigma_i^2\) fixed to its current mean (as mentioned in \eqref{eq:generalized_likelihood_full}). The resulting optimization is precisely the CMAP formulation solved by the IRLS procedure described below. The update of \(\sigma_i^2\) remains identical to that in \eqref{eq:epsilon2_update_cavi}. Thus, AMCKF preserves the core robustness and adaptivity of the rigorous CAVI framework while enabling efficient online implementation.

\subsection{The Solution to the CMAP}
\label{sec:cmap}
To solve the conditional MAP problem in~\eqref{eq:generalized_likelihood_full}, we directly minimize the MKCL with respect to the state \(x_k\). Using the whitened residuals \(\tilde{e}_k\) defined in \eqref{eq:whitened_residual}, the objective is
\[
\mathcal{J}(x_k) = \sum_{i=1}^{n+m} \rho([\tilde{e}_k]_i) = \sum_{i=1}^{n+m} \sigma_i^2 \left(1 - \exp\left(-\frac{[\tilde{e}_k]_i^2}{2\sigma_i^2}\right)\right).
\]
Taking the derivative with respect to \(x_k\) and using \(\rho'(e) = e \exp(-e^2/(2\sigma_i^2))\) gives
\[
\frac{\partial \mathcal{J}}{\partial x_k} = -\tilde{H}_k^\top W(\tilde{e}_k) \tilde{e}_k,
\]
where \(W(\tilde{e}_k) = \operatorname{diag}\left( \exp\left(-[\tilde{e}_k]_i^2 / (2\sigma_i^2)\right) \right)\). Setting this to zero yields the weighted least squares solution:
\begin{equation}
    x_k = \left( \tilde{H}_k^\top W(\tilde{e}_k) \tilde{H}_k \right)^{-1} \tilde{H}_k^\top W(\tilde{e}_k) \tilde{z}_k. \label{eq:irls_state_solution}
\end{equation}

\begin{proposition}[Monotonic Convergence of IRLS for Multi-kernel Correntropy Loss]
\label{pro:monotonic_convergence}
Let \(e(x) = z - Hx\) be a linear residual vector. Consider the objective function
\[
\mathcal{J}(x) = \sum_{i=1}^{N} \sigma_i^2 \left(1 - \exp\left(-\frac{[e(x)]_i^2}{2\sigma_i^2}\right)\right).
\]
Define the fixed-point iteration
\[
x^{(j+1)} = \left( H^\top W^{(j)} H \right)^{-1} H^\top W^{(j)} z,
\]
where \(W^{(j)} = \operatorname{diag}\left( \exp\left(-[e(x^{(j)})]_i^2/(2\sigma_i^2)\right) \right)\). Then for any starting point \(x^{(0)}\), the sequence \(\{\mathcal{J}(x^{(j)})\}_{j\ge 0}\) is monotonically non-increasing and converges to a stationary point of \(\mathcal{J}(x)\). The proof is provided in Appendix~D of the supplementary material.
\end{proposition}

\begin{remark}
Proposition~\ref{pro:monotonic_convergence} guarantees that the IRLS iteration monotonically decreases the objective function and converges to a stationary point. In practice, the iteration is initialized with the prior state estimate \(x_{k,0}^+ = \hat{x}_k^- = A \hat{x}_{k-1}\), which is optimal in the least squares sense before incorporating the current measurement.
\end{remark}

To implement the IRLS iteration efficiently, we reformulate the weighted least squares solution \eqref{eq:irls_state_solution} into a recursive Kalman-like form. Following the derivation in \cite{li2021}, denote \(x_{k,0}^+ = \hat{x}_k^-\). The state update at iteration \(t\) is written as
\begin{align}
    x_{k,t+1}^{+} &= \hat{x}_k^- + K_{k,t} \left( y_k - C \hat{x}_k^- \right), \label{eq:kf_state_update} \\
    K_{k,t} &= P_k^- C^T \left( C P_k^- C^T + \tilde{R}_{k,t} \right)^{-1}, \label{eq:kf_gain} \\
    \tilde{R}_{k,t} &= R_k^{1/2} \bar{W}(\tilde{e}_{k,t}^{\mathrm{mea}})^{-1} R_k^{1/2}, \label{eq:kf_R}
\end{align}
where \(\bar{W}(\tilde{e}_{k,t}^{\mathrm{mea}}) = \operatorname{diag}\left( \exp\left(-[\tilde{e}_{k,t}^{\mathrm{mea}}]_i^2 / (2\sigma_i^2)\right) \right)\) is the adaptive weight matrix for the measurement channels, and \(\tilde{e}_{k,t}^{\mathrm{mea}} = R_k^{-1/2}(y_k - C x_{k,t}^{+})\) is the whitened innovation.

This recursive form is computationally advantageous: the direct formulation \eqref{eq:irls_state_solution} requires inversion of the full \(n \times n\) matrix \(\tilde{H}_k^\top W(\tilde{e}_k) \tilde{H}_k\), whereas \eqref{eq:kf_gain} only requires inversion of the \(m \times m\) matrix \(C P_k^- C^T + \tilde{R}_{k,t}\), with \(m \ll n\) in typical applications. More importantly, unlike the rigorous CAVI algorithm—where the posterior covariance \(P_k^+\) must be recomputed at each iteration to evaluate the variance term \([H_k P_k^+ H_k^\top]_{ii}\) in the weight update—the AMCKF weight update depends only on the residual point estimate. The covariance update is therefore performed only once after IRLS convergence, rather than at every iteration, significantly reducing the overall computational cost.

After convergence of the IRLS iterations, the posterior error covariance is updated using the Joseph form:
\begin{equation}
    P_k^+ = (I - K_{k,t} C) P_k^- (I - K_{k,t} C)^T + K_{k,t} \tilde{R}_{k,t} K_{k,t}^T \label{eq:kf_covariance_update}
\end{equation}

The overall procedure is summarized in Algorithm~\ref{alg:cmap}.

\begin{algorithm}[!h]
\caption{Iterative CMAP State Estimator}
\label{alg:cmap}
\begin{algorithmic}[1]
\STATE \textbf{Input:} \(\hat{x}_k^-\), \(P_k^-\), \(y_k\), \(C\), \(R_k\), \(\{\sigma_i^2\}\), \(m_{\text{iter}}\), \(\epsilon\)
\STATE \textbf{Output:} \(\hat{x}_k^+\), \(P_k^{+}\)
\STATE \textbf{Initialization:} \(x_{k,0}^{+} = \hat{x}_k^-\)
\FOR{\(t = 0, 1, 2, \ldots, m_{\text{iter}}\)}
    \STATE Compute the whitened measurement innovation \(\tilde{e}_{k,t-1}^{mea} = R_k^{-1/2}(y_k - C x_{k,t}^{+})\)
    \STATE Update weights \(w_i = \exp(-[\tilde{e}_{k,t}^{mea}]_i^2 / (2\sigma_i^2))\) as in \eqref{eq:optimal_weight}
    \STATE Construct \(\bar{W} = \operatorname{diag}(w_i)\)
    \STATE Compute \(\tilde{R}_{k,t}\) as in \eqref{eq:kf_R}
    \STATE Compute Kalman gain \(K_{k,t}\) as in \eqref{eq:kf_gain}
    \STATE Update state \(x_{k,t}^{+}\) as in \eqref{eq:kf_state_update}
    \IF{\(\|x_{k,t+1}^{+} - x_{k,t}^{+}\| / \|x_{k,t+1}^{+}\| < \epsilon\)}
        \STATE break
    \ENDIF
\ENDFOR
\STATE Compute posterior covariance \(P_k^{+}\) as in \eqref{eq:kf_covariance_update}
\STATE Let \(\hat{x}_k^+ = x_{k,t}^{+}\)

\end{algorithmic}
\end{algorithm}

\subsection{Algorithm Flow of The AMCKF}
\label{sec:amckf_algorithm}

The AMCKF algorithm combines the IRLS-based CMAP state estimation from Section~\ref{sec:cmap} with the closed-form kernel width update from Section~\ref{sec:amckf_derivation}. At each time step, the state estimate \(\hat{x}_k^+\) and covariance \(P_k^+\) are first obtained via the CMAP procedure in Algorithm~\ref{alg:cmap}. Using the converged weights \(w_i\) from the CMAP iteration, the kernel width hyperparameters are then updated via \eqref{eq:epsilon2_update_cavi}, and the kernel width estimate is computed as its posterior mean \eqref{eq:sigma_posterior_mean}. The complete procedure is summarized in Algorithm~\ref{alg:mcaf}.

\begin{algorithm}[!h]
\caption{AMCKF}
\label{alg:mcaf}
\begin{algorithmic}[1]
\STATE \textbf{Input:} System matrices \(A, C\), noise covariance \(R\), initial state \(\hat{x}_{0|0}\) and \(P_{0|0}\), initial kernel widths \(\sigma_{i,0|0}^2\), initial hyperparameters \(\epsilon_1^{i,0}, \epsilon_2^{i,0}\), forgetting factor \(\rho\), filtering factor \(\beta\), active set \(\mathcal{A}\), convergence threshold \(\epsilon\), maximum iterations \(m_{\text{iter}}\).
\STATE \textbf{Output:} State estimates \(\hat{x}_k^+\) and covariance \(P_k^+\), kernel width estimates \(\hat{\sigma}_{i,k}^{2,+}\).
\FOR{each time step \(k = 1, 2, \dots\)}
    \STATE \textbf{State and Hyperparameter Prediction:}
    \STATE get \(\hat{x}_k^-\) and \(P_k^-\) via \eqref{eq:prediction_step}
    \FOR{\(i \in \mathcal{A}\)}
        \STATE get \(\epsilon_1^{i,k}\), \(\epsilon_2^{i,k}\) via \eqref{eq:hyperparameter_propagation}
    \ENDFOR
    \STATE \textbf{CMAP Iteration (State Update):}
    \STATE Perform the CMAP algorithm described in Algorithm~\ref{alg:cmap} to obtain \(\hat{x}_k^+\) and \(P_k^+\)
    \STATE \textbf{Kernel Width Update (M-step):}
    \FOR{\(i \in \mathcal{A}\)}
        \STATE Compute instantaneous update:
        \STATE get \(\epsilon_1^{i,k,+}\) as in \eqref{eq:epsilon1_update} 
        \STATE get \(\epsilon_2^{i,k,+}\) as in \eqref{eq:epsilon2_update_cavi}
        \STATE \textbf{EWMA filtering:}\\
        \(\epsilon_2^{i,k,+} = \beta \, \epsilon_2^{i,k-1,+} + (1-\beta) \, \epsilon_2^{i,k,+}\)
        \STATE Apply bounding: \(\epsilon_2^{i,k,+} \in [\epsilon_{2,\min}, \epsilon_{2,\max}]\)        
        \STATE Compute the kernel width: get \(\hat{\sigma}_{i,k}^{2,+}\) via \eqref{eq:sigma_posterior_mean}
    \ENDFOR
    \STATE \textbf{Recursion:} Set \(k \leftarrow k+1\)
\ENDFOR
\end{algorithmic}
\end{algorithm}

\section{Simulation}
\label{sec:simulation}

In this section, we evaluate the performance of the proposed AMCKF through numerical simulations. We consider a standard target tracking problem with a simulation horizon of \(20\) s and a sampling interval of 
\(\Delta t = 0.01\) s (total \(2000\) time steps). All results are 
averaged over \(100\) independent Monte Carlo runs to ensure 
statistical significance.

\subsection{Simulation Setup}

We consider a two-dimensional target tracking problem. The state vector is defined as \(x_k = [p_x, v_x, p_y, v_y]^T \in \mathbb{R}^4\), where \(p_x\) and \(p_y\) denote the positions, and \(v_x\) and \(v_y\) denote the velocities in the \(x\) and \(y\) directions, respectively. The system matrices are given by:

\begin{align}
A &= \begin{bmatrix}
1 & \Delta t & 0 & 0 \\
0 & 1 & 0 & 0 \\
0 & 0 & 1 & \Delta t \\
0 & 0 & 0 & 1
\end{bmatrix}, \quad
B = \begin{bmatrix}
\frac{1}{2}\Delta t^2 & 0 \\
\Delta t & 0 \\
0 & \frac{1}{2}\Delta t^2 \\
0 & \Delta t
\end{bmatrix}, \\[1em]
C &= \begin{bmatrix}
1 & 0 & 0 & 0 \\
0 & 0 & 1 & 0
\end{bmatrix}, \quad
D = \begin{bmatrix}
1 & 0 \\
0 & 1
\end{bmatrix},
\end{align}

\(\Delta t = 0.01\) s is the sampling interval. The process noise covariance is set to \(Q_w = I_2\) (i.e., unit covariance for the two independent acceleration noise components), and the nominal measurement noise covariance is set to \(R_v = I_2\). The initial state is \(x_0 = [0, 0, 0, 0]^T\) with initial covariance \(P_0 = 0.1 \cdot I_4\).

The simulation horizon is \(20\) s, corresponding to \(T = 2000\) time steps. All results are averaged over \(100\) independent Monte Carlo runs.

The following filters are implemented for comparison: \emph{KF}, \emph{VBKF}~\cite{sarkka2009}, \emph{STKF} (\(\nu=3\)), \emph{MKCKF} (\(\sigma^2=10\)), and the proposed \emph{AMCKF}.

For the proposed AMCKF, the initial kernel width is set to approximate 200 in every observe channel, the forgetting factor is \(\rho = 0.98\) with a threshold of \(0.99\), and the EWMA filtering factor is \(\beta = 0.1\). Physical bounds are applied with \(\epsilon_{2,{\min}} = 1\) and \(\epsilon_{2,{\max}} = 200\). The CMAP iteration stops when the relative change in state estimate falls below \(10^{-6}\) or after a maximum of \(10\) iterations.

The definitions of RMSE and the improvement rate are provided in the supplementary materials.

\subsection{Robustness Against Outliers}

To evaluate robustness, we consider three scenarios: \emph{purely Gaussian noise (Scenario 1)}, \emph{Gaussian noise with 10\% uniform outliers in \([-20,20]\) (Scenario 2)}, and \emph{Student's \(t\)-distributed noise with \(\nu=2\) (Scenario 3)}.

Table~\ref{tab:comparison} summarizes the RMSE results. Table~\ref{tab:improvement_rates} shows the improvement rates of each method relative to the standard KF across three scenarios.

Under Gaussian noise (Scenario 1), the proposed AMCKF matches the optimal performance of the standard KF, while VBKF, STKF, and MKCKF all show degraded accuracy due to their fixed robust assumptions. This demonstrates AMCKF's ability to automatically adapt to the noise environment.

Under sparse outliers (Scenario 2) and heavy-tailed noise (Scenario 3), AMCKF achieves the lowest RMSE among all methods, confirming its superior robustness. The adaptive kernel width mechanism effectively suppresses outliers while maintaining tracking capability.

Having established the advantages of AMCKF over existing methods, we next verify that it preserves the estimation accuracy of its theoretical predecessor—the rigorous CAVI algorithm.

\begin{table}[!t]
\centering
\caption{Performance comparison across three scenarios (RMSE, mean $\pm$ std).}
\label{tab:comparison}
\begin{tabular}{lccc}
\hline
\multirow{2}{*}{Method} & Scenario 1 & Scenario 2 & Scenario 3 \\
                        & (Gaussian) & (Sparse outliers) & (Student-$t$, $\nu=2$) \\
\hline
KF       & $0.1192 \pm 0.0114$ & $0.3185 \pm 0.0280$ & $0.3065 \pm 0.1386$ \\
VBKF     & $0.1951 \pm 0.0372$ & $0.3394 \pm 0.1079$ & $0.2860 \pm 0.0881$ \\
STKF     & $0.2601 \pm 0.0595$ & $0.2362 \pm 0.0594$ & $0.2637 \pm 0.0727$ \\
MKCKF    & $0.2422 \pm 0.0523$ & $0.2211 \pm 0.0536$ & $0.2521 \pm 0.0677$ \\
AMCKF & $0.1192 \pm 0.0114$ & $0.2075 \pm 0.0202$ & $0.1776 \pm 0.0174$ \\
\hline
\end{tabular}
\end{table}

\begin{table}[!t]
\centering
\caption{Improvement rates of each method relative to the standard KF across three scenarios (positive values indicate improvement).}
\label{tab:improvement_rates}
\begin{tabular}{lccc}
\hline
\multirow{2}{*}{Method} & \multicolumn{3}{c}{Improvement (\%)} \\
\cline{2-4}
                        & Scenario 1 & Scenario 2 & Scenario 3 \\
\hline
KF       & — & — & — \\
VBKF     & $-63.61\%$ & $-6.59\%$ & $+6.70\%$ \\
STKF     & $-118.18\%$ & $+25.82\%$ & $+13.98\%$ \\
MKCKF    & $-103.10\%$ & $+30.57\%$ & $+17.77\%$ \\
AMCKF & $-0.00\%$ & $+34.83\%$ & $+42.07\%$ \\
\hline
\end{tabular}
\end{table}

\subsection{Comparison with the CAVI Algorithm}
\label{sec:comparison_cavi}

To validate that the proposed AMCKF preserves the estimation accuracy of the CAVI algorithm while reducing computational cost, we compare their performance on a coupled mass-spring-damper system. The system consists of \(N=10\) masses connected by springs and dampers in a chain configuration, with both ends fixed to walls. Each mass moves in the two-dimensional plane; the state of the \(i\)-th mass is \([p_{ix}, v_{ix}, p_{iy}, v_{iy}]^\top\), yielding a total state dimension of \(40\). The measurement vector comprises the positions of all masses in both directions, resulting in a measurement dimension of \(20\). The system matrices are constructed according to the discrete-time dynamics of the coupled oscillator network, with parameters: mass \(m=1\), spring constant \(k_s=2\), damping coefficient \(c=2\), and sampling interval \(\Delta t=0.005\) s, corresponding to a damping ratio \(\zeta=c/(2\sqrt{k_s m})\approx0.707\) (underdamped). The measurement noise follows a Student-\(t\) distribution with \(\nu=3\).

\begin{figure}[!t]
\centering
\includegraphics[width=0.9\linewidth]{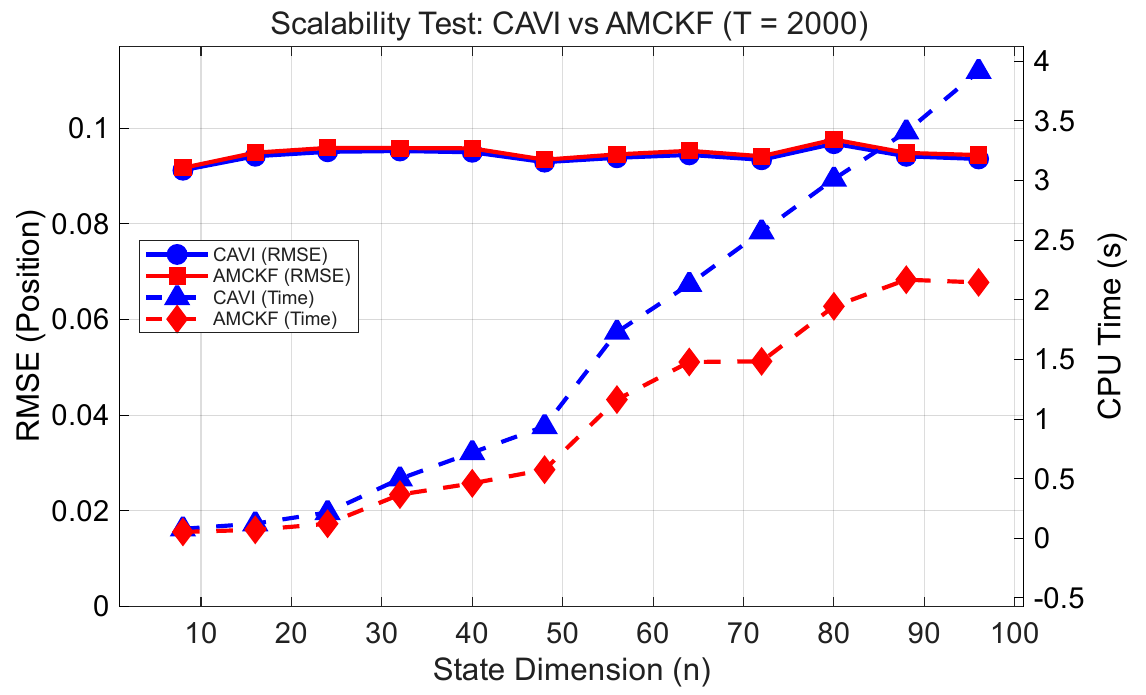}
\caption{RMSE (left axis) and CPU time (right axis) versus state dimension for CAVI and AMCKF.}
\label{fig:scalability}
\end{figure}

To examine scalability, we vary the number of masses from \(2\) to \(25\), corresponding to state dimensions from \(8\) to \(100\), while keeping all other parameters unchanged. Fig. \ref{fig:scalability} reports the RMSE and CPU time versus state dimension for both algorithms. The RMSE curves of AMCKF and CAVI remain nearly indistinguishable across all tested dimensions, confirming that the approximation does not incur significant accuracy loss regardless of problem size. Meanwhile, both algorithms exhibit quadratic growth in computation time, consistent with the \(O(n^3)\) cost of matrix inversion. AMCKF consistently reduces computational cost across all tested dimensions, with the gap widening as the state dimension increases.

\subsection{Adaptive Kernel Width Tracking}

To demonstrate the adaptive behavior of the proposed AMCKF, we design a simulation where the measurement noise environment changes over time in four consecutive 5-second stages (500 steps each at \(\Delta t = 0.01\) s): Gaussian, sparse outliers (10\% uniform in \([-100,100]\)), heavy-tailed (Student's \(t\), \(\nu=1.8\)), and return to Gaussian. The kernel width \(\sigma^2\) is recorded throughout the simulation; since \(R_v = I_2\), the estimated kernel widths for \(x\) and \(y\) are nearly identical, so only the \(x\)-direction result is shown.

\begin{figure}[!t]
\centering
\includegraphics[width=0.9\linewidth]{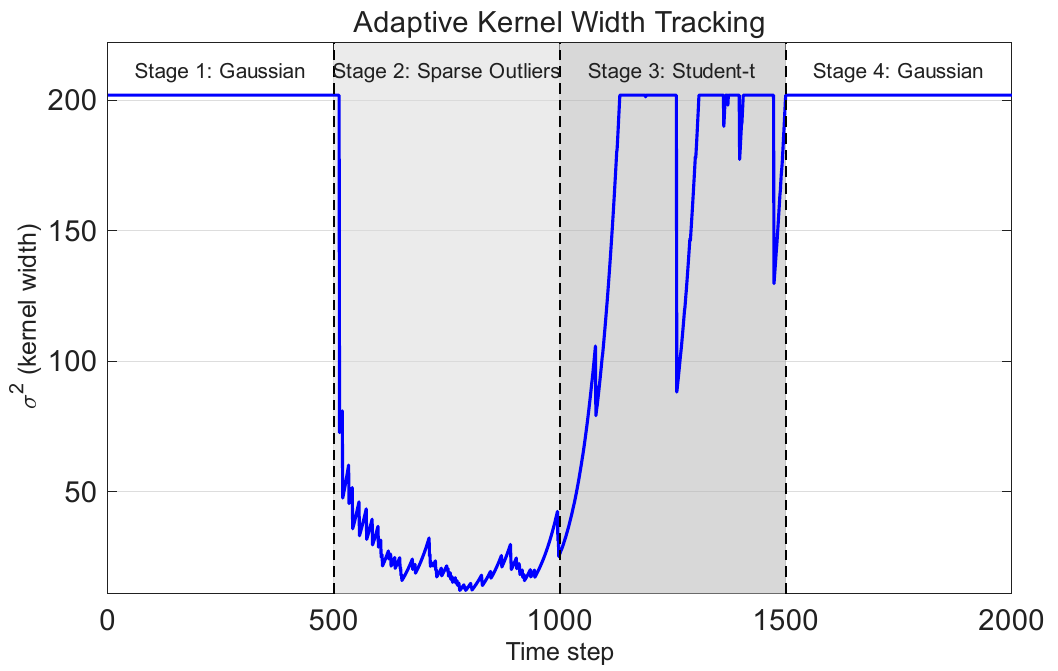}
\caption{Evolution of the estimated kernel width \(\hat{\sigma}^2\) for the \(x\)-direction measurement channel. The shaded background indicates different noise stages: white = Gaussian, light gray = sparse outliers, dark gray = heavy-tailed (Student-\(t\), \(\nu=1.8\)).}
\label{fig:sigma_tracking}
\end{figure}

Fig. \ref{fig:sigma_tracking} shows the evolution of \(\hat{\sigma}^2\). In Gaussian stages, \(\hat{\sigma}^2\) remains at its upper bound, recovering standard Kalman performance. Under outliers and heavy-tailed noise, \(\hat{\sigma}^2\) decreases to suppress corrupted measurements and recovers via EWMA filtering. The overall RMSE is \(0.1481\) (position) and \(0.1010\) (velocity), confirming that accurate tracking is maintained. These results demonstrate that AMCKF adapts its kernel width to changing noise conditions without manual tuning.

\subsection{Comprehensive Performance Comparison}

To further evaluate the proposed AMCKF, we design a mixed-challenge scenario with \(T = 2000\) time steps (\(20\) s, \(\Delta t = 0.01\) s), consisting of five consecutive stages: \emph{Gaussian} (Stage 1), \emph{sparse outliers} (Stage 2), \emph{Student-\(t\) heavy-tailed noise with \(\nu=2\)} (Stage 3), \emph{burst outliers} (Stage 4), and mixed outliers (Stage 5). All tests are conducted over 100 independent Monte Carlo runs. The proposed AMCKF is compared against standard KF, VBKF~\cite{sarkka2009}, STKF (fixed \(\nu=3\)), and MKCKF (fixed \(\sigma^2=10\)).

\begin{figure}[!t]
\centering
\includegraphics[width=0.85\linewidth]{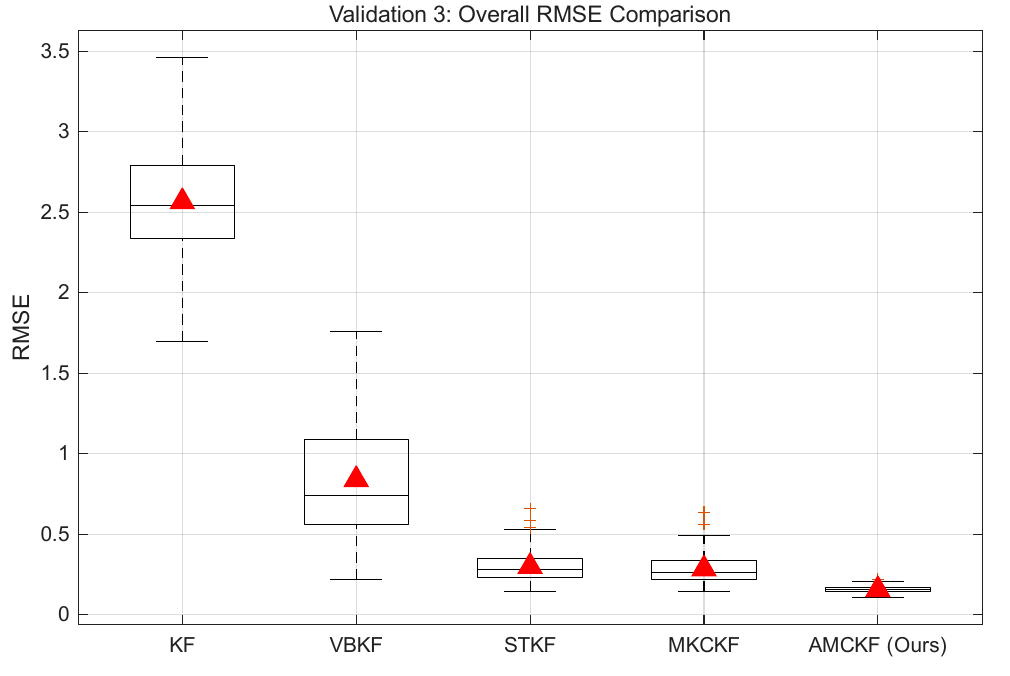}
\caption{Boxplot of overall RMSE across 100 Monte Carlo runs.
The red triangles indicate the mean values.
The proposed AMCKF achieves the lowest median and mean RMSE,
demonstrating its superior robustness and accuracy under complex noise conditions.}
\label{fig:boxplot}
\end{figure}

\begin{table}[!t]
\centering
\caption{Overall RMSE comparison (mean $\pm$ std over 100 Monte Carlo runs).}
\label{tab:overall_rmse}
\begin{tabular}{lccc}
\hline
Method & RMSE (mean $\pm$ std) & Improvement vs. KF \\
\hline
KF       & $2.5951 \pm 0.3264$ & — \\
VBKF     & $0.9396 \pm 0.4209$ & $+63.79\%$ \\
STKF     & $0.2845 \pm 0.0773$ & $+89.04\%$ \\
MKCKF    & $0.2699 \pm 0.713$ & $+89.60\%$ \\
AMCKF & $\mathbf{0.1531 \pm 0.0215}$ & $\mathbf{+94.10\%}$ \\
\hline
\end{tabular}
\end{table}

\begin{figure}[!t]
\centering
\includegraphics[width=1\linewidth]{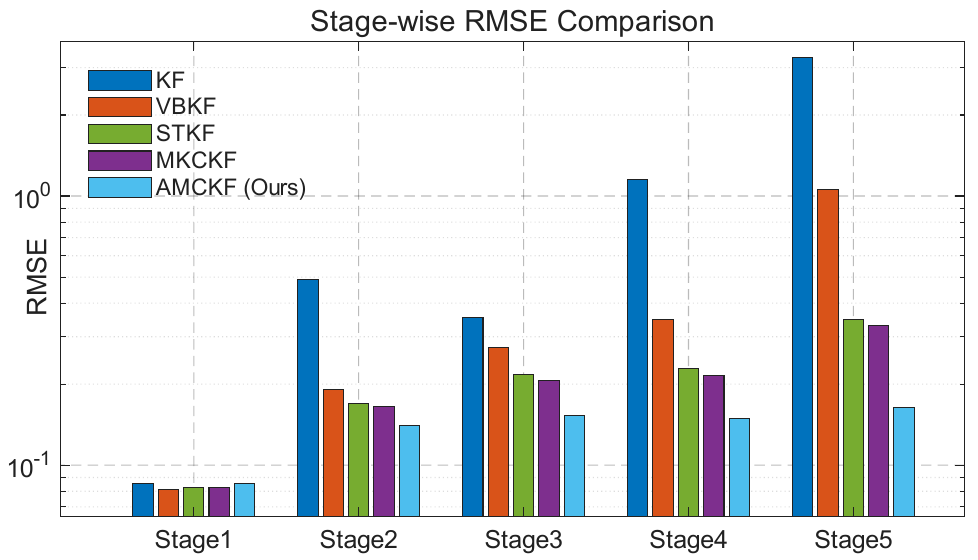}
\caption{Stage-wise RMSE comparison across five noise stages.}
\label{fig:stagewise}
\end{figure}

Fig. \ref{fig:boxplot} and Table~\ref{tab:overall_rmse} summarize the overall RMSE results. The proposed AMCKF achieves the lowest median and mean RMSE, with the smallest interquartile range among all methods, indicating superior accuracy and consistency. Fig. \ref{fig:stagewise} presents the stage-wise performance. In Stage 1 (Gaussian), all methods perform similarly. From Stage 2 onward, the AMCKF maintains a consistently low RMSE with minimal variation, while the other methods exhibit significant degradation. This stability confirms the effectiveness of the adaptive kernel width mechanism. Overall, the AMCKF effectively handles various outlier types while retaining optimal performance under Gaussian noise—a capability not available in fixed-parameter robust filters (STKF, MKCKF) or traditional adaptive filters (VBKF).

\subsection{IMU-Based Attitude Estimation on a Real Dataset: Real-World Gait Experiment}

We validate the proposed AMCKF on the Circle-Fast-Both-Disturbances gait dataset from \cite{MKCIEKFRevised} (Index 6 of it). Following the right-invariant \(\mathrm{SO}(3)\) IEKF modeling framework with decoupled magnetometer updates established in \cite{MKCIEKFRevised}. we replace the fixed kernel widths of the original MKCIEKF with the proposed adaptive variational Bayesian mechanism. The dataset features fast walking (\(\approx\)1.5 m/s) along a 2.5 m-radius circular path, with ferromagnetic plates and foot-strike vibrations jointly corrupting the IMU measurements. An Xsens MTi-670 IMU (400 Hz) is mounted on the foot, with optical motion capture providing attitude ground truth. We compare AMCKF against the commercial Xsens filter and the fixed-kernel MKCKF with a kernel width \(\sigma^2=2000\).

\begin{figure}[!t]
\centering
\includegraphics[width=0.9\linewidth]{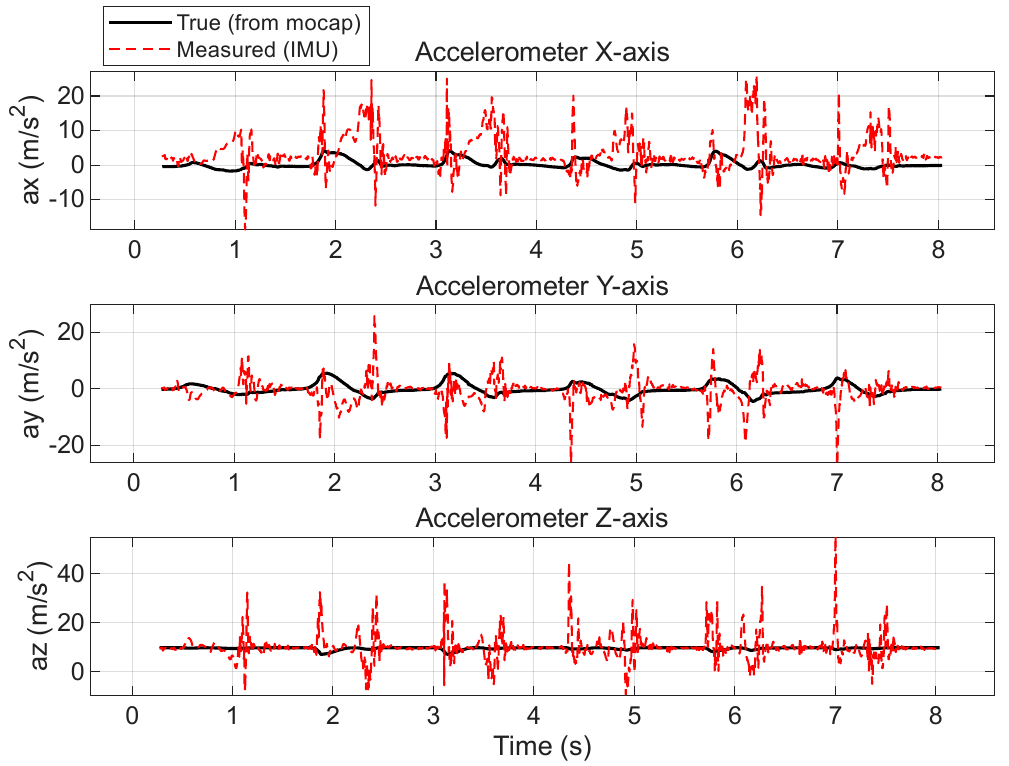}
\caption{Comparison between measured accelerometer signals (red) and true gravity (black) derived from motion capture. Deviations indicate motion‑induced acceleration during fast walking.}
\label{fig:amckf_acc_meas_vs_true}
\end{figure}

\begin{figure}[!t]
\centering
\includegraphics[width=0.9\linewidth]{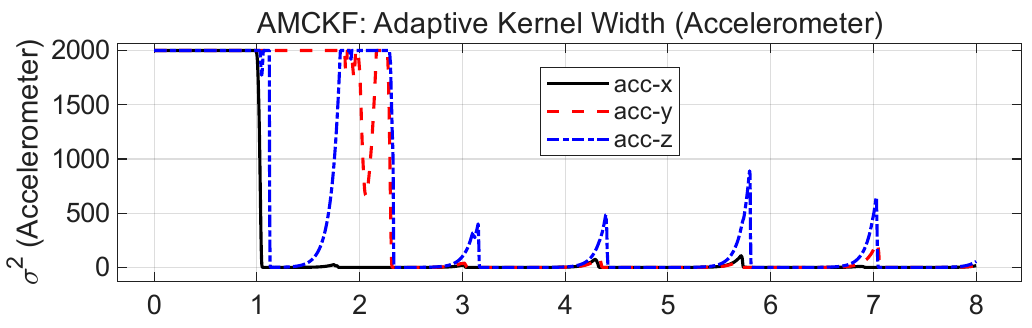}
\caption{Evolution of the adaptive kernel width \(\sigma^2\) for the accelerometer channels. The kernel width drops during strong motion‑induced acceleration and recovers afterward, showing automatic trust adjustment.}
\label{fig:amckf_acc_sigma}
\end{figure}

% Figure~\ref{fig:mag_amckf_vs_truth} compares raw magnetometer data,
% motion‑capture‑based reference magnetic field, and AMCKF‑reconstructed results.
% Raw measurements suffer severe ferromagnetic‑caused distortion,
% whereas AMCKF outputs closely follow the reference without manual tuning.

% \begin{figure}
% \centering
% \includegraphics[width=0.9\linewidth]
% {figs/mag_amckf_vs_truth.pdf}
% \caption{Comparison of magnetometer measurements in the Circle‑Fast‑Both‑Disturbances scenario. Black: reference magnetic field derived from motion capture attitude; red: raw magnetometer measurements (strongly disturbed by ferromagnetic plates); blue: magnetic field reconstructed by the proposed AMCKF. The AMCKF effectively suppresses the influence of real‑world magnetic disturbances.}
% \label{fig:mag_amckf_vs_truth}
% \end{figure}

Fig. \ref{fig:amckf_acc_meas_vs_true} compares measured acceleration with gravity from motion capture; horizontal deviations during fast walking indicate motion-induced disturbances. In response, the adaptive kernel width \(\sigma^2\) for the accelerometer channels decreases to reduce reliance on the disturbed measurements and recovers afterward, as shown in Fig.~\ref{fig:amckf_acc_sigma}.

\begin{table}[!t]
\centering
\caption{RMSE comparison on the Circle‑Fast‑Both‑Disturbances scenario (degree).}
\label{tab:amckf_circle_fast_both}
\begin{tabular}{lccc}
\hline
Method & Yaw & Roll & Pitch \\
\hline
Xsens (commercial) & 0.7111 & 0.7130 & 0.6072 \\
MKCKF & 0.6355 & 0.7269 &0.6049 \\
AMCKF (Ours) & \textbf{0.5984} & \textbf{0.7093} & \textbf{0.6071} \\
\hline
\end{tabular}
\end{table}

Table~\ref{tab:amckf_circle_fast_both} reports the Euler-angle RMSE. The fixed-kernel MKCKF with \(\sigma^2 = 2000\) performs comparably to the commercial Xsens filter. The proposed AMCKF achieves the lowest yaw error (0.5984°), outperforming both Xsens (0.7111°) and MKCKF (0.6355°), with comparable roll and pitch accuracy. This improvement demonstrates that the adaptive kernel width mechanism provides better robustness than a fixed kernel width by automatically adjusting to the observed noise statistics.

\section{Conclusion}
\label{sec:conclusion}

This paper proposed an adaptive multi-kernel correntropy Kalman filter (AMCKF) for robust state estimation. Via half-quadratic duality, we established a rigorous variational Bayesian framework with closed-form CAVI updates for joint state and kernel width estimation, and showed that the HQ-induced auxiliary variables admit degenerate variational posteriors. The AMCKF is then introduced as an efficient single-pass approximation of the exact CAVI, where the state and auxiliary variables are updated iteratively via IRLS, followed by a closed-form kernel width update. A forgetting mechanism ensures long-term adaptability. Monte Carlo simulations demonstrate that AMCKF consistently outperforms KF, VBKF, STKF, and fixed-kernel MKCKF across various outlier and heavy-tailed noise scenarios, while maintaining optimal performance under Gaussian noise and matching the accuracy of the rigorous CAVI.

Future work may address process noise adaptation, nonlinear extensions, self-tuning forgetting, theoretical analysis of kernel width dynamics.

\bibliographystyle{ieeetr}
\bibliography{ref}

\section{Supplemental Material for \\
       Adaptive Multi-kernel Correntropy Kalman Filter: 
       A Variational Bayesian Approach}

\section*{Properties of Multi-kernel Correntropy Loss}

The multi-kernel correntropy loss (MCKL) defined in (11) of the main manuscript possesses the following properties:

\begin{enumerate}
    \item \textbf{Homogeneity:} \(\rho(e) = \sigma^2 \Phi(e/\sigma)\), where \(\Phi(u) = 1 - e^{-u^2/2}\), so that \(\sigma^2\) acts purely as a scale parameter.
    
    \item \textbf{Boundedness and symmetry:} \(\rho(e)\in[0,\sigma^2)\), \(\rho(-e)=\rho(e)\).
    
    \item \textbf{Asymptotic limits:}
    \[
    \lim_{\sigma^2\to\infty} \rho(e) = \frac{e^2}{2}, \qquad
    \lim_{\sigma^2\to 0} \rho(e) = \sigma^2 \cdot \mathbf{1}_{e\neq 0}.
    \]
    
    \item \textbf{Concavity of \(\rho(\sqrt{u})\):} \(\phi(u)=\rho(\sqrt{u})\) is strictly concave on \(u\ge0\).
    
    \item \textbf{Redescending derivative:} \(\rho'(e)=e\exp(-e^2/(2\sigma^2))\), with \(\lim_{|e|\to\infty}\rho'(e)=0$.
\end{enumerate}

These properties collectively ensure that the MCKL is a bounded, symmetric, and redescending loss function. The boundedness prevents large residuals from dominating the estimation, while the redescending property ensures that the influence of extreme outliers vanishes as the residual magnitude grows. These characteristics make MCKL particularly suitable for robust estimation in the presence of heavy-tailed noise and outliers.

\section*{Performance Metrics}\label{sec:performance_metrics}

The root mean square error (RMSE) is computed separately for position and velocity in each direction:
\begin{equation}
    \text{RMSE}_{\text{pos}} = 
    \sqrt{\frac{1}{2N_{\text{MC}}T} \sum_{m=1}^{N_{\text{MC}}} \sum_{k=1}^{T} 
    \left[ (p_{x,k}^{(m)} - \hat{p}_{x,k}^{(m)})^2 + (p_{y,k}^{(m)} - \hat{p}_{y,k}^{(m)})^2 \right]},
    \label{eq:rmse_pos}
\end{equation}
and similarly for velocity:
\begin{equation}
    \text{RMSE}_{\text{vel}} = 
    \sqrt{\frac{1}{2N_{\text{MC}}T} \sum_{m=1}^{N_{\text{MC}}} \sum_{k=1}^{T} 
    \left[ (v_{x,k}^{(m)} - \hat{v}_{x,k}^{(m)})^2 + (v_{y,k}^{(m)} - \hat{v}_{y,k}^{(m)})^2 \right]},
    \label{eq:rmse_vel}
\end{equation}
where \(N_{\text{MC}} = 100\) is the number of Monte Carlo runs, \(T = 2000\) is the number of time steps, \(p_{x,k}\) and \(p_{y,k}\) denote the true positions, and \(v_{x,k}\) and \(v_{y,k}\) denote the true velocities.

The improvement rate of a given method relative to the standard Kalman filter is defined as:
\begin{equation}
    \text{Improvement} = \frac{\text{RMSE}_{\text{KF}} - \text{RMSE}_{\text{Method}}}{\text{RMSE}_{\text{KF}}} \times 100\%,
    \label{eq:improvement}
\end{equation}
where \(\text{RMSE}_{\text{KF}}\) denotes the RMSE of the standard Kalman filter, and \(\text{RMSE}_{\text{Method}}\) denotes the RMSE of the method under evaluation.

\section*{Appendix}

\subsection{Proof of Proposition~1 and robust loss functions satisfying it}
\label{appendix:proof_thm:linear_scaling}

Since \(\Phi\) satisfies the HQ conditions, by Lemma~1 there exists a function \(\psi(w)\) such that
\[
\Phi(u)=\inf_{w>0}\left[ w\frac{u^{2}}{2}+\psi(w)\right],
\]
and the infimum is attained at some \(w^{*}>0\).  
Now substitute \(u=e/\sqrt{\theta}\) into the expression for \(\rho\):
\[
\rho(e;\theta)=\theta\,\Phi\!\left(\frac{e}{\sqrt{\theta}}\right)
 =\theta\;\inf_{w>0}\left[ w\frac{(e/\sqrt{\theta})^{2}}{2}+\psi(w)\right].
\]
Because \(\theta>0\) is a constant factor, it can be multiplied into the infimum:
\[
\rho(e;\theta)=\inf_{w>0}\left[ w\frac{e^{2}}{2}+\theta\,\psi(w)\right].
\]
Thus we obtain the desired form. Note that the same \(w\) that minimizes the expression for \(\Phi\) also minimizes the transformed one (after scaling), so the range \(w>0\) is unchanged. So, \(\inf\) equals to \(\min\). This completes the proof.

Table~\ref{tab:dual_potentials} is some exemplary losses that satisfy Proposition~1. We drop the scale parameter and focus on the kernels of those losses.

\begin{table}[!h]
\centering
\caption{Representative robust loss functions satisfying Proposition~1 and their dual potential functions.}
\label{tab:dual_potentials}
\begin{tabular}{lccc}
\hline
Loss & $\Phi(u)$ & $w^*(u)$ & $\psi(w)$ \\
\hline
Welsch & $1 - e^{-u^2/2}$ & $e^{-u^2/2}$ & $w\log w - w + 1$ \\
Cauchy & $\log(1+u^2)$ & $\frac{2}{1+u^2}$ & $\log\frac{2}{w} - 1 + \frac{w}{2}$ \\
Charbonnier & $\sqrt{1+u^2}-1$ & $\frac{1}{\sqrt{1+u^2}}$ & $-\frac{1}{2}\log w + \frac{w}{2} - \frac{1}{2}$ \\
Fair & $|u| - \log(1+|u|)$ & $\frac{1}{1+|u|}$ & $\log w + \frac{1}{2w} - \frac{w}{2}$ \\
\hline
\end{tabular}
\end{table}

\subsection{Proof of Corollary~1}
\label{appendix:proof_hq_decomposition}

All conditions are satisfied according to the properties of MCKL, guaranteeing the existence of the HQ decomposition.
Applying the Half-Quadratic Duality Theorem (Lemma~1) to \(\rho(e)\) yields:
\begin{equation}
    \rho(e) = \inf_{w > 0} \left[ w\frac{e^2}{2} + \psi(w) \right].
\end{equation}
The optimal \(w\) satisfies \(w = \rho'(e) / e\). Direct computation gives \(\rho'(e) = e e^{-e^2/(2\sigma^2)}\), hence:
\begin{equation}
    w^* = \exp\left(-\frac{e^2}{2\sigma^2}\right).
\end{equation}
To determine \(\psi\), use the equality at the minimizer \(w^*\):
\begin{align}
    \rho(e) &= w^* \frac{e^2}{2} + \psi(w^*) \\
    &= -\sigma^2 w^* \ln w^* + \psi(w^*) = \sigma^2 (1 - w^*).
\end{align}
Solving for \(\psi(w^*)\) gives \(\psi(w^*) = \sigma^2 (1 - w^* + w^* \ln w^*)\). Extending to arbitrary \(w\) yields:
\begin{equation}
    \psi(w) = \sigma^2 (w \log w - w + 1).
\end{equation}

Thus, the half-quadratic decomposition of the multi-kernel correntropy loss is established:
\begin{equation}
    \rho(e) = \min_{w \in (0,1]} \left[ w \frac{e^2}{2} + \sigma^2 (w \log w - w + 1) \right].
\end{equation}
This completes the proof. 

\subsection{Proof of Theorem~1}
\label{appendix:prop:gamma_conjugacy}

For a loss function satisfying Proposition~1, the HQ-induced likelihood takes the form
\[
p(e \mid \theta, w) \propto \exp\left( -\frac{we^2}{2} - \theta\psi(w) \right).
\]
Together with the Gamma prior \(p(\theta) \propto \theta^{\epsilon_1/2} \exp(-\epsilon_2\theta/2)\), the posterior is
\[
\begin{aligned}
	&p(\theta \mid e, w) \propto \theta^{\epsilon_1/2} \exp\left( -\frac{\epsilon_2}{2}\theta \right) \cdot \exp\left( -\theta\psi(w) \right)
	\\ &= \theta^{\epsilon_1/2} \exp\left( -\left( \frac{\epsilon_2}{2} + \psi(w) \right) \theta \right).
\end{aligned}
\]
This is proportional to a Gamma distribution with shape \(\alpha = \epsilon_1/2 + 1\) and rate \(\beta = \epsilon_2/2 + \psi(w)\). Therefore, the updated hyperparameters are
\[
\epsilon_1^* = \epsilon_1, \qquad \epsilon_2^* = \epsilon_2 + 2\psi(w).
\]
Thus, the Gamma distribution is conjugate to the HQ-induced likelihood.

\subsection{Proof of Proposition~2}
\label{appendix:proof_monotonic_convergence}

Let \(e_i(x) = [z - H x]_i\) denote the \(i\)-th residual element. The objective function can be reformulated as:
\begin{equation}
    \mathcal{J}(x) = \sum_{i=1}^{N} \phi_i\left(e_i^2(x)\right),
\end{equation}
where the scalar function \(\phi_i : \mathbb{R}_{\ge 0} \rightarrow \mathbb{R}\) is defined as:
\begin{equation}
    \phi_i(u) = \sigma_i^2 \left(1 - e^{-u/(2\sigma_i^2)}\right).
\end{equation}

\subsubsection{Properties of \(\phi_i(u)\)}

For each \(i\), \(\phi_i(u)\) is continuously differentiable on \(u \ge 0\). Its first two derivatives are:
\begin{align}
    \phi_i'(u) &= \frac{1}{2} e^{-u/(2\sigma_i^2)}, \\
    \phi_i''(u) &= -\frac{1}{4\sigma_i^2} e^{-u/(2\sigma_i^2)} < 0, \quad \forall u \ge 0.
\end{align}
Since \(\phi_i''(u) < 0\) for all \(u \ge 0\), \(\phi_i(u)\) is strictly concave on \([0, \infty)\).

\subsubsection{Construction of the Surrogate Function}

For a concave function, the first-order Taylor expansion provides a global over-estimator. Specifically, for any concave function \(\phi(u)\) and any expansion point \(u_0 \ge 0\):
\begin{equation}
    \phi(u) \le \phi(u_0) + \phi'(u_0)(u - u_0).
\end{equation}

Let \(u_i^{(j)} \triangleq e_i^2(x^{(j)})\) denote the squared residual at iteration \(j\). Applying the concavity property to each term, we construct a surrogate function \(Q(x, x^{(j)})\) that serves as a global upper bound for \(\mathcal{J}(x)\):
\begin{equation}
    \mathcal{J}(x) \le Q(x, x^{(j)}) \triangleq \sum_{i=1}^{N} \left[ \phi_i(u_i^{(j)}) + \phi_i'(u_i^{(j)}) \left( e_i^2(x) - u_i^{(j)} \right) \right].
\end{equation}

\subsubsection{Minimization of the Surrogate Function}

Minimizing \(Q(x, x^{(j)})\) with respect to \(x\) is equivalent to minimizing a weighted least squares term, since the terms \(\phi_i(u_i^{(j)}) - \phi_i'(u_i^{(j)}) u_i^{(j)}\) are constant with respect to \(x\).

Define the weights:
\begin{equation}
    \omega_i^{(j)} = \phi_i'(u_i^{(j)}) = \frac{1}{2} \exp\left(-\frac{[e_i^{(j)}]^2}{2\sigma_i^2}\right).
\end{equation}
Ignoring the constant factor \(1/2\) (which can be absorbed into the weight matrix), we obtain the weight matrix:
\begin{equation}
    W^{(j)} = \operatorname{diag}\left( \exp\left(-\frac{[e_i^{(j)}]^2}{2\sigma_i^2}\right) \right).
\end{equation}

The surrogate function to be minimized is:
\begin{equation}
    Q(x, x^{(j)}) = \frac{1}{2} (z - H x)^\top W^{(j)} (z - H x) + \text{const}.
\end{equation}

This is a quadratic form in \(x\). Setting the gradient to zero yields the fixed-point iteration:
\begin{equation}
    x^{(j+1)} = \left( H^\top W^{(j)} H \right)^{-1} H^\top W^{(j)} z.
\end{equation}
This is exactly the iteration defined in Proposition~2.

\subsubsection{Monotonic Descent and Convergence}

By the majorization-minimization (MM) principle, the surrogate function \(Q(x, x^{(j)})\) satisfies:
\begin{equation}
    Q(x, x^{(j)}) \ge \mathcal{J}(x), \quad \forall x, \qquad Q(x^{(j)}, x^{(j)}) = \mathcal{J}(x^{(j)}).
\end{equation}

Since \(x^{(j+1)}\) minimizes \(Q(x, x^{(j)})\), we have:
\begin{equation}
    Q(x^{(j+1)}, x^{(j)}) \le Q(x^{(j)}, x^{(j)}) = \mathcal{J}(x^{(j)}).
\end{equation}

Combining these inequalities:
\begin{equation}
    \mathcal{J}(x^{(j+1)}) \le Q(x^{(j+1)}, x^{(j)}) \le \mathcal{J}(x^{(j)}).
\end{equation}

Thus, the sequence \(\{\mathcal{J}(x^{(j)})\}_{j \ge 0}\) is monotonically non-increasing.

\subsubsection{Convergence to a Stationary Point}

Since \(\mathcal{J}(x)\) is bounded below by zero (the multi-kernel correntropy loss is non-negative), the monotonic sequence converges to a finite limit \(\mathcal{J}^*\). Furthermore, at convergence, we have \(x^{(j+1)} = x^{(j)} \triangleq x^*\), which satisfies the stationarity condition:
\begin{equation}
    H^\top W(x^*) (z - H x^*) = 0.
\end{equation}
This is precisely the gradient condition \(\nabla \mathcal{J}(x^*) = 0\). Therefore, the limit point is a stationary point of \(\mathcal{J}(x)\). This completes the proof.

\end{document}